\documentclass[aps,pra,twocolumn,superscriptaddress,showpacs,nofootinbibfloatfix,amsmath,amsfonts,amssymb]{revtex4-2}%
\usepackage{amsmath,amsfonts,amssymb,color}
\usepackage{amsthm}
\usepackage{leftidx}
\usepackage{graphicx}
\usepackage{xcolor}
\usepackage{dcolumn}
\usepackage{bm}
\usepackage{epstopdf}
\usepackage{epsfig}
\usepackage{environ}
\usepackage{pdfcomment}

\usepackage{multirow}
\usepackage{setspace}
\usepackage{color}

\usepackage{float}
\usepackage[T1]{fontenc}
\usepackage[latin9]{inputenc}
\usepackage{setspace}
\usepackage{esint}

\providecommand{\tabularnewline}{\\}

\begin{document}


\title{Gapless higher-order topology and corner states in Floquet systems}

\author{Longwen Zhou}
\email{zhoulw13@u.nus.edu}
\affiliation{%
	College of Physics and Optoelectronic Engineering, Ocean University of China, Qingdao, China 266100
}
\affiliation{%
	Key Laboratory of Optics and Optoelectronics, Qingdao, China 266100
}
\affiliation{%
	Engineering Research Center of Advanced Marine Physical Instruments and Equipment of MOE, Qingdao, China 266100
}
\author{Rongtao Wang}
\affiliation{%
	College of Physics and Optoelectronic Engineering, Ocean University of China, Qingdao, China 266100
}
\author{Jiaxin Pan}
\affiliation{%
	College of Physics and Optoelectronic Engineering, Ocean University of China, Qingdao, China 266100
}

\date{\today}

\begin{abstract}
Higher-order topological phases (HOTPs) possess localized and symmetry-protected
eigenmodes at corners and along hinges in two and three dimensional
lattices. The numbers of these topological boundary modes will undergo
quantized changes at the critical points between different HOTPs.
In this work, we reveal unique higher-order topology induced by time-periodic
driving at the critical points of topological phase transitions, which
has no equilibrium counterparts and also goes beyond the description
of gapped topological matter. Using an alternately coupled Creutz
ladder and its Floquet-driven descendants as illustrative examples,
we analytically characterize and numerically demonstrate the zero
and $\pi$ corner modes that could emerge at the critical points between
different Floquet HOTPs. Moreover, we propose a unified scheme of
bulk-corner correspondence for both gapless and gapped Floquet HOTPs
protected by chiral symmetry in two dimensions. Our work reveals the
possibility of corner modes surviving topological transitions in Floquet
systems and initializes the study of higher-order Floquet topology
at quantum criticality.
\end{abstract}

\pacs{}
\keywords{}
\maketitle

\section{Introduction\label{sec:Int}}

Topological phases of matter constitutes one focus of research in
condensed matter physics over the past four decades. Renowned discoveries
in this area include topological insulators and topological superconductors,
whose nontrivial topology are classified by their underlying symmetries
and stabilized by the presence of spectral excitation gaps \cite{TPRev01,TPRev02,TPRev03,TPRev04,TPRev05,TPRev06,TPRev07}.
In recent years, the study of topological matter has also been driven
towards gapless systems, with the Weyl semimetal be one notable example
\cite{TMRev01,TMRev02,TMRev03,TMRev04}. Generally speaking, a gapless
symmetry-protected topological (gSPT) phase does not have an excitation
gap between its highest occupied and lowest unoccupied bulk states.
Yet, it could possess topological states that are robust to symmetry-preserving
perturbations even in the absence of a spectral gap. Compared with
topological semimetals, a generic distinction of the gSPT phase is
that it does not require any forms of gaps in the energy-momentum
parameter space to sustain its topological features. Interestingly,
it was discovered that the transition points between different topological
phases, where the spectral gaps are required to close, can also be
classified into topologically trivial and nontrivial ones \cite{gSPT01}.
The latter holds topological signatures that are usually associated
with emergent symmetries at critical points, such as degenerate zero
modes in the energy and entanglement spectra \cite{gSPT02}, whose
characterizations go beyond the standard paradigms of continuous phase
transitions and gapped topological matter \cite{gSPT03}. This intriguing
phenomenon, sometimes also referred to as topologically nontrivial
quantum critical points (QCPs) or symmetry-enriched quantum criticality,
has attracted great interest in both theoretical and experimental
studies \cite{gSPT04,gSPT05,gSPT06,gSPT07,gSPT08,gSPT09,gSPT10,gSPT11,gSPT12,gSPT13,gSPT14,gSPT15,gSPT16,gSPT17,gSPT18,gSPT19,gSPT20,gSPT21,gSPT22,gSPT23,gSPT231,gSPT24,gSPT25,gSPT26,gSPT27,gSPT28,gSPT291,gSPT292,gSPT30,gSPT31,gSPT32,gSPT33,gSPT34,gSPT35,gSPT36,gSPT37,gSPT38,gSPT39,gSPT40,gSPT41,gSPT42,gSPT43,gSPT44,gSPT45}.
Nevertheless, most existing work on topological QCPs are centered
on static one-dimensional (1D) systems in thermal equilibrium. The
possibility of having symmetry-protected gapless topology beyond equilibrium situations
and above 1D setups is much less explored.

Floquet (time-periodic) driving has been shown to be a powerful strategy
of pushing a system out-of-equilibrium while generating rich topological
phenomena with no static counterparts. For example, a high-frequency
driving field may break the time-reversal symmetry and induce the
band inversion at a Dirac cone, transforming a trivial semimetal into
a topological Chern insulator \cite{OkaPRB2009}. A near-resonant
and strong driving field may further generate long-range couplings
and greatly reshuffle the band structure of a static system, yielding
gapped Floquet phases with large topological invariants \cite{HoPRL2012}
and many topological edge states \cite{TongPRB2013}. Beyond equilibrium
analogs, time-periodic driving can also generate anomalous edge modes
with degeneracy at nonzero quasienergies \cite{JiangPRL2011} or edge
\cite{RudnerPRX2013} (and even bulk \cite{ZhouPRB2016}) bands twisting
around the whole quasienergy Brillouin zone, which could not be described
by the conventional bulk-edge correspondence of static topological
phases. These attractive phenomena have been investigated intensively
in the last fifteen years (e.g., see \cite{FTPRev01,FTPRev02,FTPRev03,FTPRev04,FTPRev05,FTPRev06,FTPRev07,FTPRev08,FTPRev09,FTPRev10}
for reviews). A key point is that the stroboscopic topological properties
of Floquet states are mainly encoded in the Floquet (one-period evolution)
operator of the system and its quasienergy (eigenphase) spectrum,
making it possible for extending the topological band theory to periodically
driven systems. With a bandstructure description, Floquet systems
offer a reasonable starting point to explore topologically nontrivial
QCPs and gSPT phases beyond equilibrium. Very recently, it was identified
that degenerate Majorana zero and $\pi$ edge modes could emerge at
the phase boundaries between different Floquet topological superconductors
\cite{FgSPT01,FgSPT02}. Moreover, the topology of these Majorana
modes \emph{cannot} be related to standard winding numbers of gapped
topological phases in one dimension. They are instead characterized
by generalized topological invariants that are tailored to work both
away from and exactly at Floquet QCPs \cite{FgSPT01,FgSPT02}. These
discoveries initialize the study of Floquet gSPT phases. Beyond one
spatial dimension, rich varieties of gapless Floquet
topology are still awaited to be revealed.

\begin{figure*}
	\begin{centering}
		\includegraphics[scale=0.5]{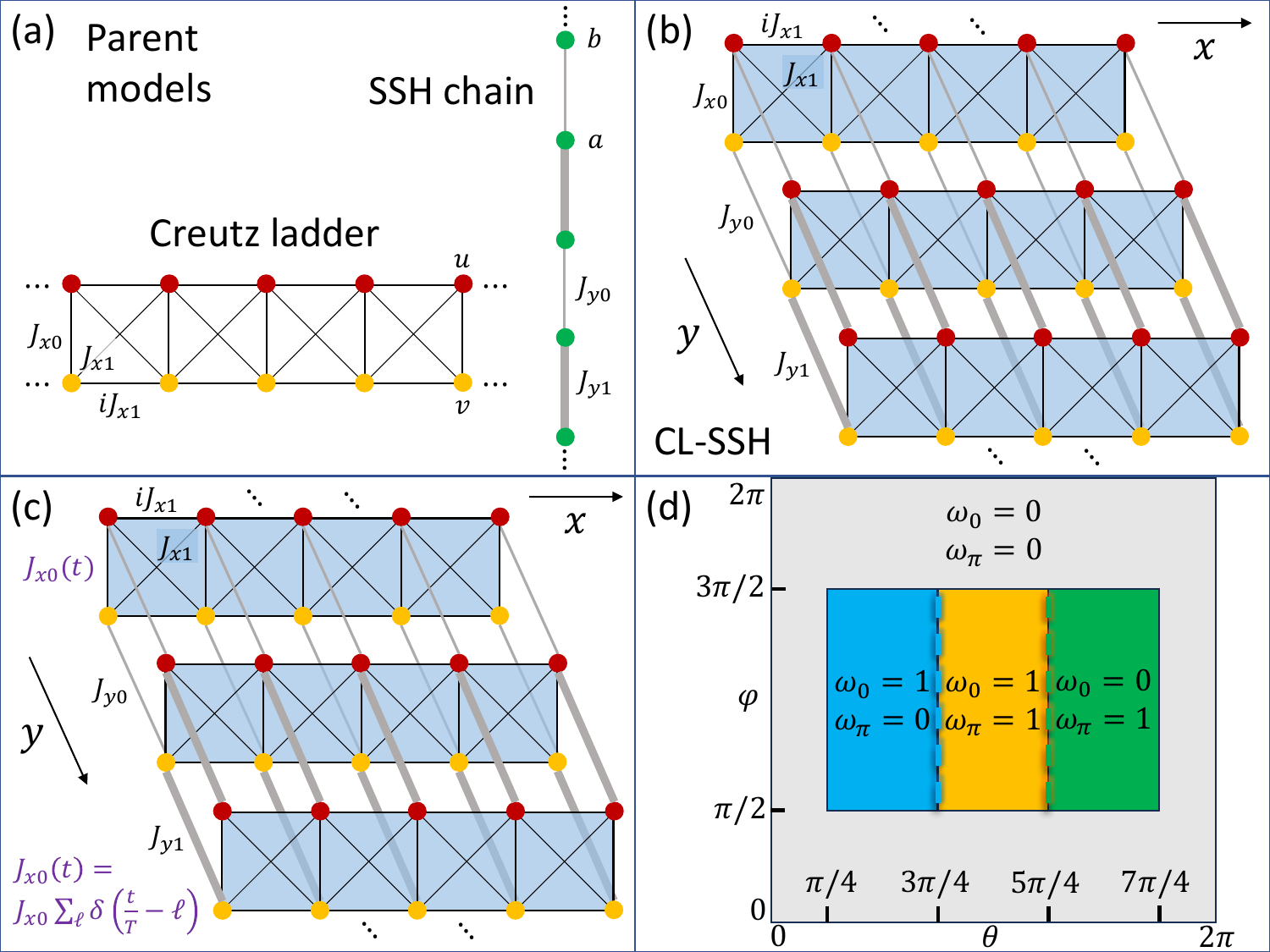}
		\par\end{centering}
	\caption{Schematic diagram of the lattice models and topological phase diagram
		of the 2D Floquet system. (a) Geometry of the 1D CL and SSH chain,
		with system parameters $(J_{x0},J_{x1},J_{y0},J_{y1})$ indexed along
		the corresponding links. (b) A series of CL, each lies along the $x$-direction
		and subject to staggered inter-ladder couplings along
		the $y$-direction, forming a 2D system that could support SOTPs.
		(c) CL-SSH coupled model under time-periodic driving, with delta kicks
		applied to all the inter-leg couplings of each CL, yielding a 2D system
		with Floquet SOTPs. (d) Phase diagram
		of the kicked CL-SSH model in (c). Each region with a
		uniform color corresponds to a Floquet SOTI phase, whose topological
		invariants $(\omega_{0},\omega_{\pi})$ are shown. The blue (green)
		dashed line denotes a topologically nontrivial phase boundary, characterized
		by the winding number $\omega_{0}=1$ ($\omega_{\pi}=1$) under PBCs
		and the number of Floquet topological corner modes $N_{0}=4$ ($N_{\pi}=4$)
		at quasienergy $0$ ($\pi$) under OBCs. Other phase boundaries are
		topologically trivial. The hopping amplitudes are parameterized as
		$J_{x0}=\frac{\pi}{2}-\frac{\pi}{4}\sin\theta$, $J_{x1}=\frac{\pi}{2}-\frac{\pi}{4}\cos\theta$,
		$J_{y0}=\frac{\pi}{2}+\frac{\pi}{4}\cos\varphi$ and $J_{y1}=\frac{\pi}{2}-\frac{\pi}{4}\cos\varphi$
		in (d). Ticks along $\theta$- and $\varphi$-axes highlight the locations
		of phase boundaries. \label{fig:Sketch}}
\end{figure*}

A topological phase of order $n$ in $d$ spatial dimensions holds
nontrivial edge states along its $(d-n)$-dimensional boundaries \cite{HOTPRev01,HOTPRev02,HOTPRev03,HOTPRev04}.
HOTPs arise in the cases with $d\geq n>1$ \cite{HOTP1,HOTP2,HOTP3,HOTP4,HOTP5,HOTP6,HOTP7,HOTP8,HOTP9},
whose Floquet counterparts were explored in a series of recent
studies \cite{FHOTP01,FHOTP02,FHOTP03,FHOTP04,FHOTP05,FHOTP06,FHOTP07,FHOTP08,FHOTP09,FHOTP10,FHOTP11,FHOTP12,FHOTP13,FHOTP14,FHOTP15,FHOTP16,FHOTP17,FHOTP18,FHOTP19,FHOTP20,FHOTP21,FHOTP22,FHOTP23,FHOTSC1,FHOTSC2,FHOTSC3,FHOTSC4,FHOTSC5,FHOTSC6,FHOTSC7,FHOTSC8,FHOTSC9,FHOTSM1,FHOTSM2,FHOTSM3,FHOTSM4,FHOTSM5}.
In this work, we uncover and characterize the nontrivial topology
and degenerate corner modes at the critical points between distinct
Floquet second-order topological phases (SOTPs) in two-dimensional
(2D) systems ($n=d=2$), thereby extending the concept of topologically
nontrivial QCPs to Floquet HOTPs. We first introduce a minimal 2D
lattice model via a coupled-ladder construction~\cite{FHOTP01} in
Sec.~\ref{sec:SOTP}. The Hamiltonian of the system possesses a second-order
topological insulator (SOTI) phase with zero-energy corner modes and
a trivial insulator phase without corner modes. We analytically show
that the critical point separating these two phases is \emph{topologically}
\emph{trivial} from the views of bulk topological indices and
corner modes. Applying time-periodic quenches to the static
model, we obtain various SOTPs unique to Floquet systems, which are
further separated by \emph{topologically nontrivial} critical points.
Theoretical and numerical descriptions of the higher-order bulk topology,
corner modes and bulk-corner correspondence emerging at these Floquet
topological transition points are systematically worked out in 
Sec.~\ref{sec:FSOTP}, which are based on generalized definitions of topological
winding numbers and analytical solutions of Floquet corner modes.
Considering different driving protocols, we demonstrate in Sec.~\ref{sec:FSOTP2}
that arbitrarily many zero- and $\pi$-quasienergy corner modes could
survive together at the critical points between different Floquet
SOTPs, providing strong evidence for the presence of nonequilibrium
higher-order topology without gap protections. In Sec.~\ref{sec:Sum},
we summarize our findings and discuss potential future studies. Further
details supplying our theoretical results are provided in Appendices~\ref{sec:App1}--\ref{sec:App5}.

\section{Static SOTPs with trivial critical points\label{sec:SOTP}}

In this section, we introduce a 2D lattice model that could
realize SOTPs through a coupled-ladder construction. The critical point separating the trivial and SOTI
phases of this model is found to be topologically trivial.

We start with a convenient scheme of generating HOTPs in $d$ spatial
dimensions, which is proceeded by coupling $n$-dimensional, lower-order
topological states along the extra $(d-n)$ dimensions ($1\leq n<d$)
\cite{FHOTP01}. One minimal example following this scheme concerns a 2D lattice model described by the Hamiltonian $H=H_{x}\oplus H_{y}$,
where $\oplus$ denotes the Kronecker sum. $H_{x}$ and $H_{y}$ represent
1D lattice Hamiltonians along the $x$ and $y$ directions of the
$x-y$ plane. The $H$ can also be written as
\begin{equation}
	H=H_{x}\otimes I_{y}+I_{x}\otimes H_{y},\label{eq:H}
\end{equation}
where $I_{x}$ ($I_{y}$) is the identity operator in the Hilbert
space of $H_{x}$ ($H_{y}$). Under the open boundary condition (OBC),
if $H_{x}$ has an edge state $|\psi\rangle$ of eigenenergy $E$
around its left boundary, while $H_{y}$ has an edge state $|\psi'\rangle$
of eigenenergy $E'$ around its lower boundary, we must have a corner
state $|\Psi\rangle=|\psi\rangle\otimes|\psi'\rangle$ of energy
$E+E'$ around the lower-left corner of the system described by
the $H$ in Eq.~(\ref{eq:H}). Moreover, if $|\psi\rangle$ ($\text{|\ensuremath{\psi'\rangle}}$)
possesses a $g$-($g'$-)fold degeneracy protected by some symmetry
${\cal G}$ (${\cal G}'$), the state $|\Psi\rangle=|\psi\rangle\otimes|\psi'\rangle$
must be $(g\cdot g')$-fold degenerate at the energy $E+E'$, which
is protected by the composite symmetry ${\cal G}\otimes{\cal G}'$
\cite{FHOTP01}. For example, let $H_{x}$ and $H_{y}$ be 1D topological
insulators with twofold degenerate edge modes at zero energy, which
maybe protected by their chiral symmetries ${\cal S}_{x}$
and ${\cal S}_{y}$. We could then have fourfold degenerate, zero-energy
corner modes in the system described by the $H$ in Eq.~(\ref{eq:H}),
which are protected by the composite chiral symmetry ${\cal S}\equiv{\cal S}_{x}\otimes{\cal S}_{y}$.
HOTPs in other spatial dimensions and with different symmetry groups
may be generated following the same strategy \cite{FHOTP01}.

In this work, we focus on the critical points of SOTPs
in 2D systems. To be concrete and without loss of generality, we choose
the $H_{x}$ and $H_{y}$ in Eq.~(\ref{eq:H}) as Hamiltonians
of the Creutz ladder (CL) model \cite{Creutz1999PRL} and the Su-Schrieffer-Heeger
(SSH) model~\cite{SSH1979PRL}, whose lattice geometries are
illustrated in Fig.~\ref{fig:Sketch}(a).
We emphasize that the general theoretical formulas we obtained regarding topological
invariants and bulk-corner correspondence in this work do not rely on the specific choice of these models.
Our only general requirement is that both the $H_{x}$ and $H_{y}$ should be two-band Hamiltonians with chiral symmetries.
At half-filling, both
models constitute prototypes of 1D topological insulators with symmetry-protected
edge zero modes. Under periodic boundary conditions (PBCs), their
Bloch Hamiltonians in momentum space are given by
\begin{alignat}{1}
	H_{x}(k_{x}) & =[J_{x0}+J_{x1}\cos(k_{x})]\sigma_{x}+J_{x1}\sin(k_{x})\sigma_{z},\label{eq:Hxkx}\\
	H_{y}(k_{y}) & =[J_{y0}+J_{y1}\cos(k_{y})]\tau_{x}+J_{y1}\sin(k_{y})\tau_{y},\label{eq:Hyky}
\end{alignat}
where $\sigma_{x,y,z}$ and $\tau_{x,y,z}$ are Pauli matrices. 
We will use $\sigma_{0}$ and $\tau_{0}$
to denote $2\times2$ identity matrices. $k_{x}\in[-\pi,\pi)$ and
$k_{y}\in[-\pi,\pi)$ are quasimomenta along the $x$ and $y$ directions.
In Eq.~(\ref{eq:Hxkx}), $J_{x0}\in\mathbb{R}$ denotes the nearest-neighbor
(NN) intracell coupling between the two legs of the CL, while the
NN intercell coupling along each leg and between the two legs are
set to be the same as $J_{x1}\in\mathbb{R}$. There is a $\pi$ magnetic
flux through each plaquette of the CL \cite{Creutz1999PRL}.
In Eq.~(\ref{eq:Hyky}), $J_{y0}\in\mathbb{R}$ and $J_{y1}\in\mathbb{R}$
denote the intracell and intercell NN hopping amplitudes along the
SSH chain. The characterization of topological phases in CL and SSH models are
well-established (e.g., see Ref.~\cite{Basu2024Book} for a review).
Below we recap them briefly.

The Hamiltonian $H_{x}(k_{x})$ possesses the chiral symmetry
${\cal S}_{x}=\sigma_{y}$, the time-reversal symmetry ${\cal T}_{x}=\sigma_{x}$
and the particle-hole symmetry ${\cal C}_{x}=\sigma_{z}$, in the
sense that 
\begin{equation}
	\begin{cases}
		{\cal S}_{x}H_{x}(k_{x}){\cal S}_{x}=-H_{x}(k_{x}), & \qquad{\cal S}_{x}^{2}=\sigma_{0},\\
		{\cal T}_{x}H_{x}^{*}(k_{x}){\cal T}_{x}^{\dagger}=H_{x}(-k_{x}), & \qquad{\cal T}_{x}^{2}=\sigma_{0},\\
		{\cal C}_{x}H_{x}^{*}(k_{x}){\cal C}_{x}^{\dagger}=-H_{x}(-k_{x}), & \qquad{\cal C}_{x}^{2}=\sigma_{0}.
	\end{cases}\label{eq:HxSym}
\end{equation}
Therefore, the CL in Eq.~(\ref{eq:Hxkx}) belongs
to the symmetry class BDI \cite{TPRev04}. At half-filling, its gapped
topological phases are characterized by a winding number
$w_{x}$, i.e.,
\begin{equation}
	w_{x}=\int_{-\pi}^{\pi}\frac{dk_{x}}{2\pi}\frac{\partial\phi_{x}(k_{x})}{\partial k_{x}},\label{eq:wx}
\end{equation}
where $\phi_{x}(k_{x})=\arctan\{J_{x1}\sin(k_{x})/[J_{x0}+J_{x1}\cos(k_{x})]\}$.
$w_{x}$ counts the number of times that the two-component vector
$(J_{x0}+J_{x1}\cos(k_{x}),J_{x1}\sin(k_{x}))$ winds around the origin
when $k_{x}$ goes over the first Brillouin zone. It can be inspected
from Eqs.~(\ref{eq:Hxkx}) and (\ref{eq:wx}) that
\begin{equation}
	w_{x}=\begin{cases}
		0, & |J_{x1}|<|J_{x0}|,\\
		\pm1/2, & |J_{x1}|=|J_{x0}|,\\
		\pm1, & |J_{x1}|>|J_{x0}|.
	\end{cases}\label{eq:wx2}
\end{equation}
Therefore, at half-filling, the CL belongs to a topologically nontrivial
(trivial) insulating phase with the winding number $w_{x}=\pm1$ ($w_{x}=0$).
Under the OBC, two degenerate edge modes at zero energy will appear
in the topologically nontrivial phase due to the bulk-edge
correspondence $N_{0x}=2|w_{x}|$, where $N_{0x}$ counts the number
of edge zero modes. At the critical point $|J_{x1}|=|J_{x0}|$, the
system undergoes a topological transition accompanied by the
bulk-gap closing of $H_{x}(k_{x})$ at zero energy. We obtain a half-integer-quantized
$w_{x}$ at the critical point due to Eq.~(\ref{eq:wx}), which
could not tell us whether the phase transition point is topologically
trivial \cite{gSPT18}. To overcome this issue, we follow
Ref.~\cite{gSPT02} to find an analytic continuation
of $H_{x}(k_{x})$, yielding the complex mapping function
\begin{equation}
	f_{x}(z)=J_{x0}+J_{x1}z,\qquad z\in\mathbb{C}.\label{eq:fxz}
\end{equation}
Inside the unit circle $|z|=1$, the number of zeros $N_{xz}$ of
$f_{x}(z)$ minus the number of its poles $N_{xp}$ (including their
multiplicities and orders) defines a generalized invariant
\begin{equation}
	\omega_{x}\equiv N_{xz}-N_{xp}.\label{eq:omex}
\end{equation}
For the CL, a direct inspection of Eq.~(\ref{eq:fxz}) yields
\begin{equation}
	\omega_{x}=\begin{cases}
		0, & |J_{x1}|\leq|J_{x0}|,\\
		1, & |J_{x1}|>|J_{x0}|.
	\end{cases}\label{eq:omex2}
\end{equation}
Therefore, we have $\omega_{x}=w_{x}$ in the gapped phases of CL.
The critical point $|J_{x1}|=|J_{x0}|$ tends out to be topologically
trivial with $\omega_{x}=0$, having no degenerate edge modes at zero
energy under the OBC. The topological triviality of this critical
point can be verified by computing the zero-energy solution of the
CL under the OBC. Taking a half-infinite chain with unit-cell indices
$m=1,2,...,\infty$, we find the zero-energy solution as
\begin{equation}
	|\psi_{{\rm L}}\rangle=\sum_{m=1}^{\infty}\left(-\frac{J_{x0}}{J_{x1}}\right)^{m-1}(\hat{u}_{m}^{\dagger}-i\hat{v}_{m}^{\dagger})|\emptyset\rangle,\label{eq:psi0}
\end{equation}
where ``L'' means left, $\hat{u}_{m}^{\dagger}$ ($\hat{v}_{m}^{\dagger}$)
creates a fermion in the sublattice $u$ ($v$) of the $m$th unit
cell on the upper (lower) leg of the CL {[}see Fig.~\ref{fig:Sketch}(a){]}.
$|\emptyset\rangle$ denotes the vacuum state. There would be
another edge mode if the OBC is taken at $m=\infty$ (see Appendix~\ref{sec:App1} for details). These edge modes become
extended at $|J_{x1}|=|J_{x0}|$, so that they could not survive at
the critical point. We then propose a refined bulk-edge correspondence
for the CL,
\begin{equation}
	N_{0x}=2\omega_{x}=\begin{cases}
		0, & |J_{x1}|\leq|J_{x0}|,\\
		2, & |J_{x1}|>|J_{x0}|,
	\end{cases}\label{eq:CLBBC}
\end{equation}
which holds true in both gapped phases and along phase boundaries.
It confirms that the transition point between topological and
trivial insulator phases of the CL is itself topologically trivial.

The symmetry, topology and bulk-edge correspondence of the SSH model
can be treated similarly. We first identify the chiral
symmetry ${\cal S}_{y}=\tau_{z}$, time-reversal symmetry ${\cal T}_{y}=\tau_{0}$,
and particle-hole symmetry ${\cal C}_{y}=\tau_{z}$ of the $H_{y}(k_{y})$
in Eq.~(\ref{eq:Hyky}), i.e.,
\begin{equation}
	\begin{cases}
		{\cal S}_{y}H_{y}(k_{y}){\cal S}_{y}=-H_{y}(k_{y}), & \qquad{\cal S}_{y}^{2}=\tau_{0},\\
		{\cal T}_{y}H_{y}^{*}(k_{y}){\cal T}_{y}^{\dagger}=H_{y}(-k_{y}), & \qquad{\cal T}_{y}^{2}=\tau_{0},\\
		{\cal C}_{y}H_{y}^{*}(k_{y}){\cal C}_{y}^{\dagger}=-H_{y}(-k_{y}), & \qquad{\cal C}_{y}^{2}=\tau_{0}.
	\end{cases}\label{eq:HySym}
\end{equation}
The SSH model thus also belongs to the symmetry class BDI \cite{TPRev04}.
At half-filling, its gapped topological phases can be characterized
by a winding number $w_{y}\in\mathbb{Z}$, i.e.,
\begin{equation}
	w_{y}=\int_{-\pi}^{\pi}\frac{dk_{y}}{2\pi}\frac{\partial\phi_{y}(k_{y})}{\partial k_{y}},\label{eq:wy}
\end{equation}
where $\phi(k_{y})=\arctan\{J_{y1}\sin(k_{y})/[J_{y0}+J_{y1}\cos(k_{y})]\}$.
$w_{y}$ counts the number of times that the two-component vector
$(J_{y0}+J_{y1}\cos(k_{y}),J_{y1}\sin(k_{y}))$ encircles the origin
when $k_{y}$ goes over the first Brillouin zone. It can be deduced
from Eqs.~(\ref{eq:Hyky}) and (\ref{eq:wy}) that
\begin{equation}
	w_{y}=\begin{cases}
		0, & |J_{y1}|<|J_{y0}|,\\
		\pm1/2, & |J_{y1}|=|J_{y0}|,\\
		\pm1, & |J_{y1}|>|J_{y0}|.
	\end{cases}\label{eq:wy2}
\end{equation}
Following the approach of
Ref.~\cite{gSPT02}, the complex mapping function of $H_{y}(k_{y})$
should take the form
\begin{equation}
	f_{y}(z)=J_{y0}+J_{y1}z,\qquad z\in\mathbb{C}.\label{eq:fyz}
\end{equation}
Counting the zeros $N_{yz}$ and poles $N_{yp}$ of $f_{y}(z)$ inside
the unit circle $|z|<1$ yields the generalized winding
number for the SSH model as
\begin{equation}
	\omega_{y}\equiv N_{yz}-N_{yp}=\begin{cases}
		0, & |J_{y1}|\leq|J_{y0}|,\\
		1, & |J_{y1}|>|J_{y0}|.
	\end{cases}\label{eq:omey}
\end{equation}
It further predicts the bulk-edge correspondence
\begin{equation}
	N_{0y}=2\omega_{y}=\begin{cases}
		0, & |J_{y1}|\leq|J_{y0}|,\\
		2, & |J_{y1}|>|J_{y0}|,
	\end{cases}\label{eq:SSHBBC}
\end{equation}
where $N_{0y}$ counts the number of degenerate zero-energy edge modes
under OBCs at both edges of the chain. In conclusion,
we find that when the intracell (intercell) hopping is stronger, the
ground-state of the SSH model at half-filling is a topologically
trivial (nontrivial) insulating phase, which has no (two) degenerate
edge modes at zero energy under the OBC. The critical point $|J_{y1}|=|J_{y0}|$
should be topologically trivial too, as it has $\omega_{y}=0$ and
there are no degenerate edge zero modes ($N_{0y}=0$). This
triviality can be verified by computing zero-energy solutions
of the SSH model. Considering a half-infinite chain with unit-cell
indices $n=1,2,...,\infty$, the zero-energy solution
reads
\begin{equation}
	|\psi'_{{\rm B}}\rangle=\sum_{n=1}^{\infty}\left(-\frac{J_{y0}}{J_{y1}}\right)^{n-1}\hat{a}_{n}^{\dagger}|\emptyset\rangle,\label{eq:varphi0}
\end{equation}
where ``B'' means bottom, $\hat{a}_{n}^{\dagger}$ creates a fermion
in the sublattice $a$ of the $n$th unit cell {[}see Fig.~\ref{fig:Sketch}(a){]},
and $|\emptyset\rangle$ denotes the vacuum state. There is one more
edge zero mode if the OBC is taken at $n=\infty$ (see
Appendix \ref{sec:App1} for details). These edge modes could not survive at $|J_{y1}|=|J_{y0}|$,
where they become extended into the bulk. They can only be exponentially
localized when $|J_{y1}|>|J_{y0}|$, i.e., within the topologically
nontrivial gapped phase. Therefore, the transition point between the topological
and trivial phases of the SSH model is indeed topologically
trivial. In parallel with the CL, Eqs.~(\ref{eq:omey}) and (\ref{eq:SSHBBC})
depict the topology and bulk-edge correspondence of the SSH model
throughout its parameter space, including both gapped and gapless
regions.

\begin{table*}
	\caption{Summary of the results for the CL, SSH and coupled CL-SSH models regarding
		their bulk topological phases, phase boundaries (critical
		points) and bulk-boundary (-edge or -corner) correspondence. Meanings
		of the system parameters are given in the paragraph below Eqs.~(\ref{eq:Hxkx})
		and (\ref{eq:Hyky}). Definitions of the winding numbers are given
		in Eqs.~(\ref{eq:omex}), (\ref{eq:omey}) and (\ref{eq:ome}). $N_{0x}$
		and $N_{0y}$ are the numbers of degenerate edge modes of the CL and
		SSH models at zero energy, respectively. $N_{0}$ is the number of
		degenerate corner modes of the coupled CL-SSH model at zero energy
		under OBCs. \label{tab:1}}
	\begin{centering}
		\begin{tabular}{|c|c|c|c|}
			\hline 
			\multirow{2}{*}{Model} & Topological & Topological & Bulk-boundary\tabularnewline
			& phase boundary & invariant & correspondence\tabularnewline
			\hline 
			\hline 
			CL & $|J_{x1}|=|J_{x0}|$ & $\omega_{x}=\left\{ \begin{array}{cc}
				0, & |J_{x1}|\leq|J_{x0}|\\
				1, & |J_{x1}|>|J_{x0}|
			\end{array}\right.$ & $N_{0x}=2\omega_{x}$\tabularnewline
			\hline 
			SSH & $|J_{y1}|=|J_{y0}|$ & $\omega_{y}=\left\{ \begin{array}{cc}
				0, & |J_{y1}|\leq|J_{y0}|\\
				1, & |J_{y1}|>|J_{y0}|
			\end{array}\right.$ & $N_{0y}=2\omega_{y}$\tabularnewline
			\hline 
			Coupled & $|J_{x1}|=|J_{x0}|$ or & \multirow{2}{*}{$\omega=\left\{ \begin{array}{cc}
					0, & |J_{x1}|\leq|J_{x0}|\,\,{\rm or}\,\,|J_{y1}|\leq|J_{y0}|\\
					1, & |J_{x1}|>|J_{x0}|\,\,\&\,\,|J_{y1}|>|J_{y0}|
				\end{array}\right.$} & \multirow{2}{*}{$N_{0}=4\omega$}\tabularnewline
			CL-SSH & $|J_{y1}|=|J_{y0}|$ &  & \tabularnewline
			\hline 
		\end{tabular}
		\par\end{centering}
\end{table*}

With the CL and SSH models as building blocks, we could
construct a ``minimal'' model holding 2D SOTPs, whose Hamiltonian
under the PBC takes the form
\begin{equation}
	H(k_{x},k_{y})=H_{x}(k_{x})\otimes\tau_{0}+\sigma_{0}\otimes H_{y}(k_{y}),\label{eq:Hkxky}
\end{equation}
where $H_{x}(k_{x})$ and $H_{y}(k_{y})$ are given by the Hamiltonians
of CL and SSH models in Eqs.~(\ref{eq:Hxkx}) and (\ref{eq:Hyky}).
The lattice geometry of this system is shown
in Fig.~\ref{fig:Sketch}(b). As discussed before, if the 1D CL and SSH parents both possess zero-energy edge
modes, their 2D descendant in Eq.~(\ref{eq:H}) will hold
zero modes around the corners of the 2D lattice under
the OBCs when their protecting symmetries (here the chiral symmetries)
are preserved. To be explicit, consider a semi-infinite system with
unit-cell indices $(m,n)\in\mathbb{Z}^{+}\times\mathbb{Z}^{+}$, which
has an open corner at $(m,n)=(1,1)$. With Eqs.~(\ref{eq:psi0}) and
(\ref{eq:varphi0}), we find a zero-energy (un-normalized) solution of the 2D
system as
\begin{widetext}
\begin{equation}
	|\Psi_{{\rm LB}}\rangle\equiv|\psi_{{\rm L}}\rangle\otimes|\psi'_{{\rm B}}\rangle=\sum_{m,n=1}^{\infty}\left(-\frac{J_{x0}}{J_{x1}}\right)^{m-1}\left(-\frac{J_{y0}}{J_{y1}}\right)^{n-1}(|m,u\rangle-i|m,v\rangle)\otimes|n,a\rangle.\label{eq:Phi}
\end{equation}
\end{widetext}
This state is exponentially localized around $(m,n)=(1,1)$ (i.e.,
a corner zero mode) if and only if $|J_{x1}|>|J_{x0}|$ and $|J_{y1}|>|J_{y0}|$,
implying that both the CL and SSH models must be set in their topologically
nontrivial gapped regions. Based on Eqs.~(\ref{eq:omex}) and (\ref{eq:omey}),
a composite invariant $\omega$ can be introduced to account
this fact, i.e., 
\begin{equation}
	\omega\equiv\omega_{x}\omega_{y}=\begin{cases}
		1, & |J_{x1}|>|J_{x0}|\,\,\&\,\,|J_{y1}|>|J_{y0}|,\\
		0, & {\rm otherwise}.
	\end{cases}\label{eq:ome}
\end{equation}
At half-filling, our 2D system then represents an SOTI (a trivial
insulator) if $\omega=1$ ($\omega=0$). The integer-quantization
of $\omega$ is preserved so long as the composite chiral symmetry
${\cal S}\equiv{\cal S}_{x}\otimes{\cal S}_{y}=\sigma_{y}\otimes\tau_{z}$
of $H(k_{x},k_{y})$ {[}Eq.~(\ref{eq:Hkxky}){]}, satisfying ${\cal S}H(k_{x},k_{y}){\cal S}=-H(k_{x},k_{y})$
and ${\cal S}^{2}=\sigma_{0}\otimes\tau_{0}$
remains intact. If $|\Psi_{{\rm LB}}\rangle$
represents a corner zero mode, we will have three other corner zero
modes under the OBCs along
both dimensions (see Appendix~\ref{sec:App1} for details). The total number of these corner modes $N_{0}$ can
thus be associated to the winding number $\omega$ as
\begin{equation}
	N_{0}=4\omega=\begin{cases}
		4, & |J_{x1}|>|J_{x0}|\,\,\&\,\,|J_{y1}|>|J_{y0}|,\\
		0, & {\rm otherwise}.
	\end{cases}\label{eq:BBC}
\end{equation}
Notably, all the corner modes disappear in the critical parameter
space $|J_{x1}|=|J_{x0}|$ with any $(J_{y0},J_{y1})$, or $|J_{y1}|=|J_{y0}|$
with any $(J_{x0},J_{x1})$. The topological phase boundaries of our
2D coupled CL-SSH model are thus topologically trivial, characterized
by the vanishing of generalized invariant $\omega$ and the delocalization
of corner zero mode $|\Psi_{{\rm LB}}\rangle$ in Eq.~(\ref{eq:Phi}).

In Table \ref{tab:1}, we summarize the main results regarding the
topology and bulk-edge correspondence of the CL, SSH and coupled
CL-SSH models. The key point is that at the critical
points between topological and trivial gapped phases, symmetry-protected degenerate edge or corner zero modes are
\emph{absent} in these models. In this sense, the phase boundaries between distinct
insulating phases in these static systems are topologically trivial.
In the next section, we will show that under Floquet driving, the
original critical points of our static 2D system could become topologically
nontrivial, carrying fourfold-degenerate Floquet eigenmodes with zero
and $\pi$ quasienergies at corners. These higher-order topological
states can coexist with a gapless quasienergy bulk, making them unique
to Floquet critical systems. Moreover, new phase boundaries holding
topological corner modes can be generated by Floquet driving, which
have no counterparts in the static setup.

\section{Floquet SOTPs with topologically nontrivial critical points\label{sec:FSOTP}}
We now apply Floquet driving to the 2D lattice model introduced in
the last section. To emphasize the main physics, we focus on a simple
driving protocol, in which the intracell coupling between the two
legs of each CL is subject to a string of equally paced delta pulses
{[}see Fig.~\ref{fig:Sketch}(c) for an illustration{]}. The time-dependent
Bloch Hamiltonian of such a delta-kicked CL-SSH model reads
\begin{equation}
	H(k_{x},k_{y},t)=H_{x}(k_{x},t)\otimes\tau_{0}+\sigma_{0}\otimes H_{y}(k_{y}),\label{eq:Hkt}
\end{equation}
where $H_{y}(k_{y})$ is the Hamiltonian of SSH model in Eq.~(\ref{eq:Hyky}),
\begin{alignat}{1}
	H_{x}(k_{x},t)& = H_{x0}\delta_{T}(t)+H_{x1}(k_{x}),\label{eq:Hxkxt}\\
	H_{x0}& = J_{x0}\sigma_{x},\label{eq:Hx0t}\\
	H_{x1}(k_{x})& = J_{x1}[\cos(k_{x})\sigma_{x}+\sin(k_{x})\sigma_{z}],\label{eq:Hx1kx}
\end{alignat}
$\delta_{T}(t)\equiv\sum_{\ell\in\mathbb{Z}}\delta(t/T-\ell)$.
The NN inter-leg couplings are thus only turned on in a narrow time
window at the intersection of each two adjacent periods $\ell T$
and $(\ell+1)T$. The Floquet operator of the system in $(k_{x},k_{y})$-space,
defined by its evolution operator $U(k_{x},k_{y})=\hat{\mathsf{T}}e^{-\frac{i}{\hbar}\int_{0}^{T}H(k_{x},k_{y},t)dt}$
over a complete driving period $T$ reads
\begin{equation}
	U(k_{x},k_{y})=e^{-iH_{x0}}e^{-iH_{x1}(k_{x})}\otimes e^{-iH_{y}(k_{y})},\label{eq:Uk}
\end{equation}
where we have set $\hbar/T=1$ as the unit of energy and $\hat{\mathsf{T}}$
performs the time-ordering. The Floquet operator $U(k_{x},k_{y})$
has a direct product structure, which allows us to analyze it by considering
its 1D parent systems 
\begin{alignat}{1}
	U_{x}(k_{x}) & \equiv e^{-iH_{x0}}e^{-iH_{x1}(k_{x})},\label{eq:Uxkx}\\
	U_{y}(k_{y}) & \equiv e^{-iH_{y}(k_{y})}.\label{eq:Uyky}
\end{alignat}

First, we notice that the effective Hamiltonian of $U_{y}(k_{y})$,
defined as $i\ln U_{y}(k_{y})$, is just the static Hamiltonian
$H_{y}(k_{y})$ of SSH model {[}Eq.~(\ref{eq:Hyky}){]} mod $2\pi$,
whose symmetry, topology and bulk-edge correspondence have be analyzed
in the last section. The quasienergy spectrum of $U_{y}(k_{y})$ is
given by $\pm\varepsilon'(k_{y})$ mod $2\pi$, with 
\begin{equation}
	\varepsilon'(k_{y})=\sqrt{[J_{y0}+J_{y1}\cos(k_{y})]^{2}+[J_{y1}\sin(k_{y})]^{2}}.
\end{equation}
At half-filling, the gapped topological phases of $U_{y}(k_{y})$
are also characterized by the winding number $w_{y}$ in Eq.~(\ref{eq:wy}).
The bulk-edge correspondence throughout its parameter space (including
both gapped and gapless regions) can be further captured by the generalized
invariant $\omega_{y}$ in Eq.~(\ref{eq:omey}).

The Floquet-induced nontrivial topology of $U(k_{x},k_{y})$ should
then be controlled mainly by its parent $U_{x}(k_{x})$, which describes
a periodically kicked CL \cite{ZhouPRA2020}. The quasienergy spectrum
of $U_{x}(k_{x})$ can be obtained by solving its eigenvalue equation
$U_{x}(k_{x})|\psi\rangle=e^{-i\varepsilon(k_{x})}|\psi\rangle$,
yielding the Floquet bands $\pm\varepsilon(k_{x})$ mod $2\pi$ with
\begin{equation}
	\varepsilon(k_{x})=\arccos(\cos J_{x0}\cos J_{x1}-\sin J_{x0}\sin J_{x1}\cos k_{x}).\label{eq:Ekx}
\end{equation}
The $\pm\varepsilon(k_{x})$ could touch with each other at
either the center ($\varepsilon=0$) or the boundary $(\varepsilon=\pi)$
of the first quasienergy Brillouin zone $\varepsilon\in[-\pi,\pi]$.
In the parameter space $(J_{x0},J_{x1})$, this allows us to determine
the gap-closing condition between $\pm\varepsilon(k_{x})$, i.e.,
\begin{equation}
	J_{x0}\pm J_{x1}=\nu\pi,\qquad\nu\in\mathbb{Z}.\label{eq:PB}
\end{equation}
In Eq.~(\ref{eq:PB}), we take the $+$ ($-$) sign if the critical
quasimomentum of gap-closing point appears at $k_{x}=0$ ($k_{x}=\pi$).
The two bands meet at $\varepsilon=0$ ($\varepsilon=\pi$)
when $\nu$ takes even (odd) integers. In comparison, the static CL
{[}Eq.~(\ref{eq:Hxkx}){]} only allows $\nu=0$ (so that $|J_{x1}|=|J_{x0}|$)
in order for the system to be critical (gapless). We conclude that
the addition of a simple Floquet driving may indeed enrich the phase
structure of the static CL. This enrichment can be further
carried over to the 2D composite system $U(k_{x},k_{y})=U_{x}(k_{x})\otimes U_{y}(k_{y})$
following Eq.~(\ref{eq:Uk}), as will be revealed shortly.

The $U_{x}(k_{x})$ in Eq.~(\ref{eq:Uxkx}) does not show any explicit
symmetries like the static CL in Eq.~(\ref{eq:Hxkx}).
To figure out the symmetry-protected topological properties of $U_{x}(k_{x})$,
we introduce unitary transformations to a pair of symmetric time frames
\cite{STF}, in which the $U_{x}(k_{x})$ reads
\begin{alignat}{1}
	U_{1x}(k_{x})= & e^{-\frac{i}{2}H_{x1}(k_{x})}e^{-iH_{x0}}e^{-\frac{i}{2}H_{x1}(k_{x})},\label{eq:U1xkx}\\
	U_{2x}(k_{x})= & e^{-\frac{i}{2}H_{x0}}e^{-H_{x1}(k_{x})}e^{-\frac{i}{2}H_{x0}}.\label{eq:U2xkx}
\end{alignat}
The $U_{\alpha x}(k_{x})$ ($\alpha=1,2$) and $U_{x}(k_{x})$ are
unitary equivalent, implying that they have the same quasienergy spectrum.
Meanwhile, $U_{\alpha x}(k_{x})$ has the chiral, time-reversal and
particle-hole symmetries ${\cal S}_{x}=\sigma_{y}$, ${\cal T}_{x}=\sigma_{x}$
and ${\cal C}_{x}=\sigma_{z}$, in the sense that
\begin{equation}
	\begin{cases}
		{\cal S}_{x}U_{\alpha x}(k_{x}){\cal S}_{x}=U_{\alpha x}^{\dagger}(k_{x}), & \qquad{\cal S}_{x}^{2}=\sigma_{0},\\
		{\cal T}_{x}U_{\alpha x}^{*}(k_{x}){\cal T}_{x}^{\dagger}=U_{\alpha x}^{\dagger}(-k_{x}), & \qquad{\cal T}_{x}^{2}=\sigma_{0},\\
		{\cal C}_{x}U_{\alpha x}^{*}(k_{x}){\cal C}_{x}^{\dagger}=U_{\alpha x}(-k_{x}), & \qquad{\cal C}_{x}^{2}=\sigma_{0},
	\end{cases}\label{eq:UxSym}
\end{equation}
where $\alpha=1,2$. Therefore, the system described
by $U_{x}(k_{x})$ also belongs to the symmetry class BDI. Its gapped
topological phases can be characterized by a pair of
winding numbers $(w_{0},w_{\pi})$ \cite{STF}. To define
these topological invariants, we first obtain the winding number of
the effective Hamiltonian of $U_{\alpha x}(k_{x})$ as
\begin{equation}
	w_{\alpha x}=\int_{-\pi}^{\pi}\frac{dk_{x}}{2\pi}\frac{\partial\phi_{\alpha x}(k_{x})}{\partial k_{x}},\label{eq:w12x}
\end{equation}
where $\phi_{\alpha x}(k_{x})\equiv\arctan(d_{z\alpha}/d_{x\alpha})$,
$d_{x\alpha}=\frac{i}{2}{\rm Tr}[\sigma_{x}U_{\alpha x}(k_{x})]$
and $d_{z\alpha}=\frac{i}{2}{\rm Tr}[\sigma_{z}U_{\alpha x}(k_{x})]$
for $\alpha=1,2$. Combining the $(w_{1x},w_{2x})$
allows us to define topological invariants for
the kicked CL {[}Eq.~(\ref{eq:Uxkx}){]} as
\begin{equation}
	(w_{0x},w_{\pi x})=\frac{1}{2}(w_{1x}+w_{2x},w_{1x}-w_{2x}).\label{eq:w0px}
\end{equation}
For 1D Floquet systems in the BDI symmetry class, it has been shown
that the winding numbers $(w_{0x},w_{\pi x})$, as defined in 
Eq.~(\ref{eq:w0px}), could fully characterize the topological properties
and bulk-edge correspondences of their gapped phases
\cite{STF}. Specially, if $N_{0x}$ and $N_{\pi x}$ denote the numbers
of degenerate edge modes at zero and $\pi$ quasienergies in a gapped phase, we would
have
\begin{equation}
	(N_{0x},N_{\pi x})=2(|w_{0x}|,|w_{\pi x}|).\label{eq:N0px}
\end{equation}

\begin{table*}
	\caption{Summary of the key results for the kicked CL-SSH model concerning
		its Floquet SOTPs, phase boundaries (i.e., critical points), and bulk-corner
		correspondence. Definition of the winding numbers are given in 
		Eq.~(\ref{eq:ome0p}). $N_{0}$ and $N_{\pi}$ are the numbers of degenerate
		corner modes at zero and $\pi$ quasienergies under OBCs, respectively.
		\label{tab:2}}
	\begin{centering}
		\begin{tabular}{|c|c|}
			\hline 
			Model & Kicked CL-SSH\tabularnewline
			\hline 
			\hline 
			Topological & $J_{x0}\pm J_{x1}=\nu\pi,\qquad\nu\in\mathbb{Z}$\tabularnewline
			phase boundary & or $|J_{y1}|=|J_{y0}|$\tabularnewline
			\hline 
			Topological  & $(1,1),$ $\,\,|J_{y1}|>|J_{y0}|\,\&\,\left|\tan\frac{J_{x1}}{2}\right|>\left|\tan\frac{J_{x0}}{2}\right|\,\&\,\left|\tan\frac{J_{x0}}{2}\tan\frac{J_{x1}}{2}\right|>1$\tabularnewline
			\cline{2-2} 
			invariants & $(1,0)$, $\,\,|J_{y1}|>|J_{y0}|\,\&\,\left|\tan\frac{J_{x1}}{2}\right|>\left|\tan\frac{J_{x0}}{2}\right|\,\&\,\left|\tan\frac{J_{x0}}{2}\tan\frac{J_{x1}}{2}\right|\leq1$\tabularnewline
			\cline{2-2} 
			\multicolumn{1}{|c|}{$(\omega_{0},\omega_{\pi})=$} & $(0,1)$, $\,\,|J_{y1}|>|J_{y0}|\,\&\,\left|\tan\frac{J_{x1}}{2}\right|\le\left|\tan\frac{J_{x0}}{2}\right|\,\&\,\left|\tan\frac{J_{x0}}{2}\tan\frac{J_{x1}}{2}\right|>1$\tabularnewline
			\cline{2-2} 
			& $(0,0)$, $\,\,$otherwise\tabularnewline
			\hline 
			Bulk-corner  & \multirow{2}{*}{$(N_{0},N_{\pi})=4(\omega_{0},\omega_{\pi})$}\tabularnewline
			correspondence & \tabularnewline
			\hline 
		\end{tabular}
		\par\end{centering}
\end{table*}

Along the critical lines {[}Eq.~(\ref{eq:PB}){]}, the winding numbers
$(w_{1x},w_{2x})$ may take integer or half-integer values,
making the bulk-edge correspondence in Eq.~(\ref{eq:N0px}) inapplicable.
We can resolve this issue by generalizing the definition of winding
numbers in accordance with Eq.~(\ref{eq:omex}) and computing the
edge-state solutions analytically at the critical points of the kicked
CL. In short, we introduce a pair of generalized winding numbers $(\omega_{x1},\omega_{x2})$
for $U_{1x}(k_{x})$ and $U_{2x}(k_{x})$ (see Appendix
\ref{sec:App2} for details). Their linear combinations yield the new topological
invariants
\begin{equation}
	(\omega_{x0},\omega_{x\pi})=\frac{1}{2}(\omega_{x1}+\omega_{x2},\omega_{x1}-\omega_{x2}).\label{eq:omex0p12}
\end{equation}
These invariants are both integer-quantized, and they are applicable
in both the gapped phases and at the critical points of the kicked
CL (or any other 1D chiral-symmetric driven systems with two bands).
In terms of these generalized invariants, the bulk-edge correspondence
can be re-established throughout the phase diagram (see Fig.~\ref{fig:KCLPD}
in Appendix \ref{sec:App2}) as
\begin{equation}
	(N_{0x},N_{\pi x})=2(|\omega_{0x}|,|\omega_{\pi x}|).\label{eq:KCLBBC}
\end{equation}
This relation is confirmed by the exact solutions Floquet edge modes
at quasienergies zero and $\pi$ in both gapped phases and along topological
phase boundaries of the kicked CL (see Appendix \ref{sec:App2} for details).

We are now ready to unveil the Floquet-induced higher-order topology
in the gapped and gapless phases of our 2D system $U(k_{x},k_{y})$.
Starting with Eq.~(\ref{eq:Uk}) and incorporating the transformations
in Eqs.~(\ref{eq:U1xkx}) and (\ref{eq:U2xkx}), we arrive at the
Floquet operators of our kicked CL-SSH coupled model in two symmetric
time frames, i.e.,
\begin{equation}
	U_{\alpha}(k_{x},k_{y})=U_{\alpha x}(k_{x})\otimes e^{-iH_{y}(k_{y})},\quad\alpha=1,2.\label{eq:U12kxy}
\end{equation}
From Eqs.~(\ref{eq:HySym}) and (\ref{eq:UxSym}), we can
identify the chiral, time-reversal and particle-hole symmetries of
$U_{\alpha}(k_{x},k_{y})$, i.e., ${\cal S}=\sigma_{y}\otimes\tau_{z}$,
${\cal T}=\sigma_{x}\otimes\tau_{0}$ and ${\cal C}=\sigma_{z}\otimes\tau_{z}$.
Under their operations, the $U_{\alpha}(k_{x},k_{y})$ ($\alpha=1,2$)
transforms as
\begin{equation}
	\left\{ \begin{array}{c}
		{\cal S}U_{\alpha}(k_{x},k_{y}){\cal S}=U_{\alpha}^{\dagger}(k_{x},k_{y}),\\
		{\cal T}U_{\alpha}^{*}(k_{x},k_{y}){\cal T}^{\dagger}=U_{\alpha}^{\dagger}(-k_{x},-k_{y}),\\
		{\cal C}U_{\alpha}^{*}(k_{x},k_{y}){\cal C}^{\dagger}=U_{\alpha}(-k_{x},-k_{y}),
	\end{array}\right.\label{eq:Usym}
\end{equation}
where ${\cal S}^{2}={\cal T}^{2}={\cal C}^{2}=1$. Our kicked CL-SSH
model thus belongs to the symmetry class BDI, whose first-order
Floquet topological phases are all trivial in two dimensions. Nevertheless,
the driven system described by Eq.~(\ref{eq:Uk}) could possess
Floquet SOTPs. To characterize their topology and bulk-corner correspondence,
we propose the generalized winding numbers $(\omega_{0},\omega_{\pi})$
as
\begin{equation}
	(\omega_{0},\omega_{\pi})\equiv(|\omega_{0x}\omega_{y}|,|\omega_{\pi x}\omega_{y}|).\label{eq:ome0p}
\end{equation}
Here, $(\omega_{0x},\omega_{\pi x})$ are topological invariants of
the kicked CL in Eq.~(\ref{eq:omex0p12}), which can also topologically
characterize other 1D, two-band chiral-symmetric driven systems. The
$\omega_{y}$, as defined in Eq.~(\ref{eq:omey}), is just the generalized
winding number of SSH model. Using these bulk topological invariants,
we can further count the numbers of Floquet zero and $\pi$ corner
modes $N_{0}$ and $N_{\pi}$ under the OBCs through the relation
\begin{equation}
	(N_{0},N_{\pi})=4(\omega_{0},\omega_{\pi}).\label{eq:CLSSHBBC}
\end{equation}
As a main result of this study, the rule of bulk-corner correspondence
in Eq.~(\ref{eq:CLSSHBBC}) not only works in all the gapped phases
of our kicked CL-SSH model, but also holds along its topological phase
boundaries, where at least one spectral gap of $U(k_{x},k_{y})$ at
the quasienergies zero and $\pi$ closes. Eq.~(\ref{eq:CLSSHBBC})
thus captures the gapped and gapless higher-order topology of our
kicked CL-SSH model from both the bulk and corner perspectives. In
Appendices \ref{sec:App3} and \ref{sec:App4}, we justify the 
Eq.~(\ref{eq:CLSSHBBC}) for our system by evaluating both the $(N_{0},N_{\pi})$
and $(\omega_{0},\omega_{\pi})$ analytically, with the key results
summarized in Table \ref{tab:2}. We emphasize that the validity of
Eq.~(\ref{eq:CLSSHBBC}) is not restricted to our kicked CL-SSH setup.
This correspondence is satisfied by any 2D, chiral-symmetric Floquet
system whose Hamiltonian shares the form of Eq.~(\ref{eq:Hkt}), in
which the driving field is only applied along
one spatial dimension.

The topological phase diagram of our kicked CL-SSH model is reported
in Fig.~\ref{fig:Sketch}(d). In parameter regions where the spectra
of $U(k_{x},k_{y})$ are gapped at both zero and $\pi$ quasienergies,
we find one trivial insulator phase with $(\omega_{0},\omega_{\pi})=(0,0)$
and three other Floquet SOTI phases. While the SOTI phase with $(\omega_{0},\omega_{\pi})=(1,0)$
can appear in static systems, the phases with $(\omega_{0},\omega_{\pi})=(0,1)$
and $(\omega_{0},\omega_{\pi})=(1,1)$ are new and unique to Floquet
settings, with each of them carrying four degenerate corner modes
at the quasienergy $\pi$ under OBCs. Besides, the phase with $(\omega_{0},\omega_{\pi})=(1,1)$
possesses four other Floquet corner modes at quasienergy zero. Importantly,
we identify two topologically nontrivial phase boundaries, as highlighted
by the vertical dashed lines in Fig.~\ref{fig:Sketch}(d), whose topological
properties are endowed uniquely by Floquet driving fields. Along the
critical line with $\varphi\in(\pi/2,3\pi/2)$ at $\theta=3\pi/4$
in Fig.~\ref{fig:Sketch}(d), we have four Floquet corner modes at
zero quasienergy, which are coexistent with a bulk spectrum that is
gapless at quasienergy $\pi$. The number of these corner modes is
counted by the topological invariants $(\omega_{0},\omega_{\pi})=(1,0)$
according to Eq.~(\ref{eq:CLSSHBBC}). Along the critical line with
$\varphi\in(\pi/2,3\pi/2)$ at $\theta=5\pi/4$ in Fig.~\ref{fig:Sketch}(d),
we have four Floquet corner modes at $\pi$ quasienergy, which are
coexistent with a bulk spectrum that is gapless at quasienergy zero.
These corner $\pi$ modes are unique to Floquet systems and their
number is counted by the topological invariant $\omega_{\pi}=1$ due
to Eq.~(\ref{eq:CLSSHBBC}). Notably, all the topologically nontrivial
phase boundaries will vanish when the driving field is switched off.
They are thus of Floquet-origin and represent (together with their
associated corner modes) what we do mean by gapless higher-order topology
in Floquet systems.

\begin{figure}
	\begin{centering}
		\includegraphics[scale=0.485]{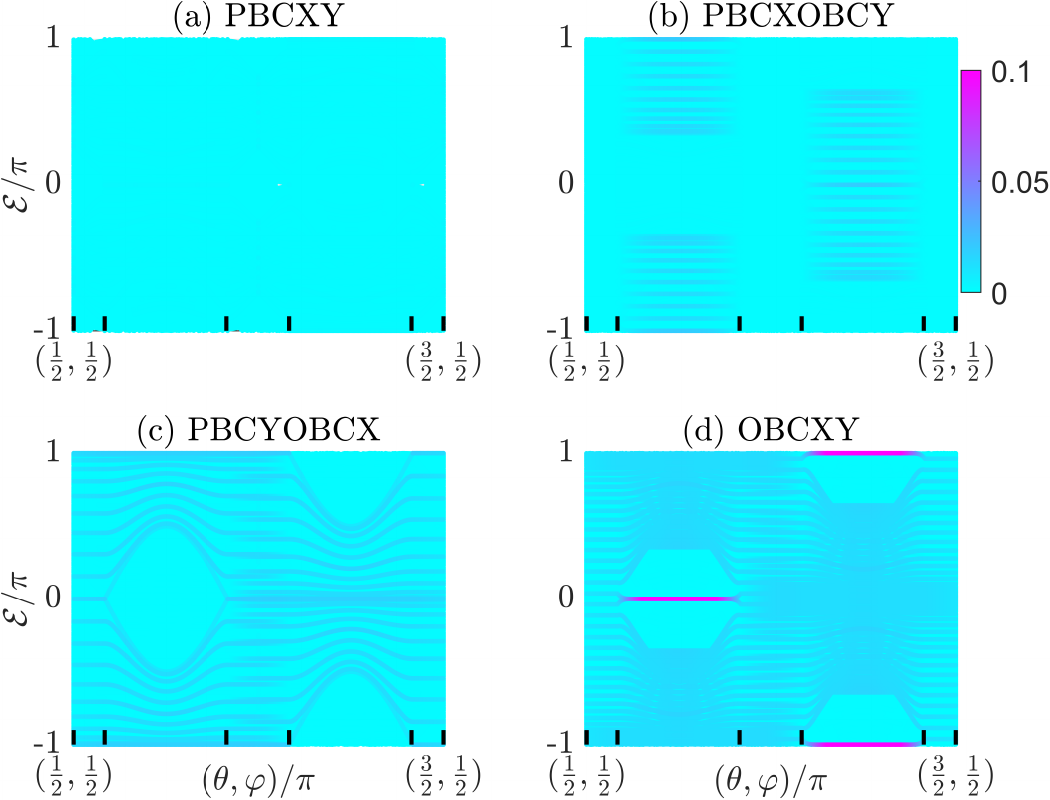}
		\par\end{centering}
	\caption{Quasienergy spectra ${\cal E}$ of the kicked CL-SSH model evaluated
		along topological phase boundaries. The boundary conditions of the
		system are (a) PBCs along both $x$ and $y$ directions, (b) PBC along
		$x$ and OBC along $y$ directions, (c) PBC along $y$ and OBC along
		$x$ directions, and (d) OBCs along both $x$ and $y$ directions.
		The variables $(\theta,\varphi)$ and related system parameters $(J_{x0},J_{x1},J_{y0},J_{y1})$
		are defined following Fig.~\ref{fig:Sketch}(d). The ticks along the
		horizontal axis of each panel denote $(\theta,\varphi)=(\pi/2,\pi/2)$,
		$(3\pi/4,\pi/2)$, $(3\pi/4,3\pi/2)$, $(5\pi/4,3\pi/2)$, $(5\pi/4,\pi/2)$
		and $(3\pi/2,\pi/2)$ from left to right. Adjacent ticks are connected
		by straight lines in the $(\theta,\varphi)$-space of Fig.~\ref{fig:Sketch}(d),
		yielding a trajectory along the boundaries between different Floquet
		phases. The color of each data point represents the IPR of the associated
		eigenstate, whose value is given by the shared color bar of (a)--(d).
		The system size is $L_x=L_y=40$ for all panels.
		\label{fig:CLSSH1}}
\end{figure}

In the rest of this section, we present numerical results to support
our theoretically findings. We first consider the Floquet spectra
of kicked CL-SSH model under different boundary conditions, which
are computed along topological phase boundaries and presented in 
Fig.~\ref{fig:CLSSH1}. The quasienergy spectrum can be obtained by first
taking the Fourier transformation of the Floquet operator in Eq.~(\ref{eq:Uk})
from momentum to position representations, and then solving the eigenvalue
equation $U|\Psi\rangle=e^{-i{\cal E}}|\Psi\rangle$ under given boundary
conditions. In our calculation, we take the unit of energy as the
ratio between Planck constant $\hbar$ and driving period $T$, i.e.,
$\hbar/T=1$. The ${\cal E}$ is then defined mod $2\pi$ and its
first quasienergy Brillouin zone is $[-\pi,\pi]$. The Floquet eigenstate
can be expanded in position representation $\{|n_{x},n_{y}\rangle\}$
as $|\Psi\rangle=\sum_{n_{x}=1}^{L_{x}}\sum_{n_{y}=1}^{L_{y}}\Psi_{n_{x},n_{y}}|n_{x},n_{y}\rangle$,
where $L_{x}$ and $L_{y}$ denote the number of lattice sites along
$x$ and $y$ directions. The probability distribution of $|\Psi\rangle$
in lattice space is given by $|\Psi_{n_{x},n_{y}}|^{2}$ at different
sites $(n_{x},n_{y})$. The inverse participation ratio (IPR) of $|\Psi\rangle$
is defined as ${\rm IPR}=\sum_{n_{x},n_{y}}|\Psi_{n_{x},n_{y}}|^{4}$,
which goes to zero (remains finite) in the limits $L_{x}\rightarrow\infty$
and $L_{y}\rightarrow\infty$ if $|\Psi\rangle$ represents an extended
(a localized) state.

The calculations in Fig.~\ref{fig:CLSSH1} are all performed via exact diagonalization in the lattice representation. In each panel of Fig.~\ref{fig:CLSSH1}, the horizontal axis corresponds to the system parameters $(\theta,\varphi)$. Each point along the horizontal axis gives a value of $(\theta,\varphi)$ taken along the topological phase boundary of Fig.~\ref{fig:Sketch}(d). At each $(\theta,\varphi)$, we plot all the quasienergy eigenvalues of our system along the vertical axis under the boundary condition listed in the corresponding figure caption. For our 2D system of lattice size $40\times40$, there are $1600$ such eigenvalues at each $(\theta,\varphi)$, which fill the first quasienergy Brillouin zone almost uniformly. The color of each data point gives the IPR of corresponding eigenstate, which goes to zero (one) for a bulk (corner) state in the limit of large system size. The magnitude of IPR is given by the shared color bar of all figure panels.

In Figs.~\ref{fig:CLSSH1}(a)--\ref{fig:CLSSH1}(d), we find that
under any boundary conditions, there are no observable quasienergy
gaps in ${\cal E}\in[-\pi,\pi]$ throughout the considered parameter
space along topological phase boundaries. This is expected, as the
quasienergy bands should meet with each other when the system parameters
are tuned to the critical points between different Floquet topological
phases. Remarkably, we find exponentially localized eigenmodes only
when the OBCs are taken along both $x$ and $y$ directions, as shown in Fig.~\ref{fig:CLSSH1}(d). These modes appear
at either the quasienergy zero or $\pi$, with their numbers $N_{0}=4$
or $N_{\pi}=4$ predicted exactly by the bulk-corner correspondence
in Eq.~(\ref{eq:CLSSHBBC}) and the solutions of Floquet corner modes
in Appendix \ref{sec:App3}. Moreover, they only appear in parameter
domains with $(\omega_{0},\omega_{\pi})=(1,0)$ and $(\omega_{0},\omega_{\pi})=(0,1)$
in Table \ref{tab:2}. We thus verified that corner-localized
zero and $\pi$ eigenmodes, as defining features of Floquet SOTPs,
could indeed appear at phase boundaries and resist topological phase
transitions, forming gapless higher-order topological states at critical
points. Specially, the gapless higher-order topology associated to
degenerate $\pi$ corner modes represents a nonequilibrium criticality
unique to Floquet-driven systems. Referring to Fig.~\ref{fig:Sketch}(d),
we realize that a topologically nontrivial phase boundary (critical
points with corner modes) could only be sandwiched between \emph{topologically
	distinct and nontrivial} gapped phases. This explains why we find
no signatures of corner modes and higher-order topology at any critical
points in the non-driven CL-SSH model {[}Eq.~(\ref{eq:Hkxky}){]},
as the phase boundaries there are all between trivial 
and higher-order topological phases. Besides, we notice in Fig.~\ref{fig:CLSSH1}(d)
that with the change of system parameters along topological phase
boundaries, transitions between gapless Floquet SOTPs (with zero or
$\pi$ corner modes) and gapless trivial phases (without corner modes)
could happen. These transitions can be regarded as ``\emph{topological
	phase transitions of topological phase transitions}'', which are
also characterized by the topological invariants $(\omega_{0},\omega_{\pi})$
in Eq.~(\ref{eq:ome0p}) and the bulk-corner correspondence in 
Eq.~(\ref{eq:CLSSHBBC}). The numerical results in 
Fig.~\ref{fig:CLSSH1} are all consistent with our theoretical predictions
(as summarized in Table \ref{tab:2}).

\begin{figure}
	\begin{centering}
		\includegraphics[scale=0.482]{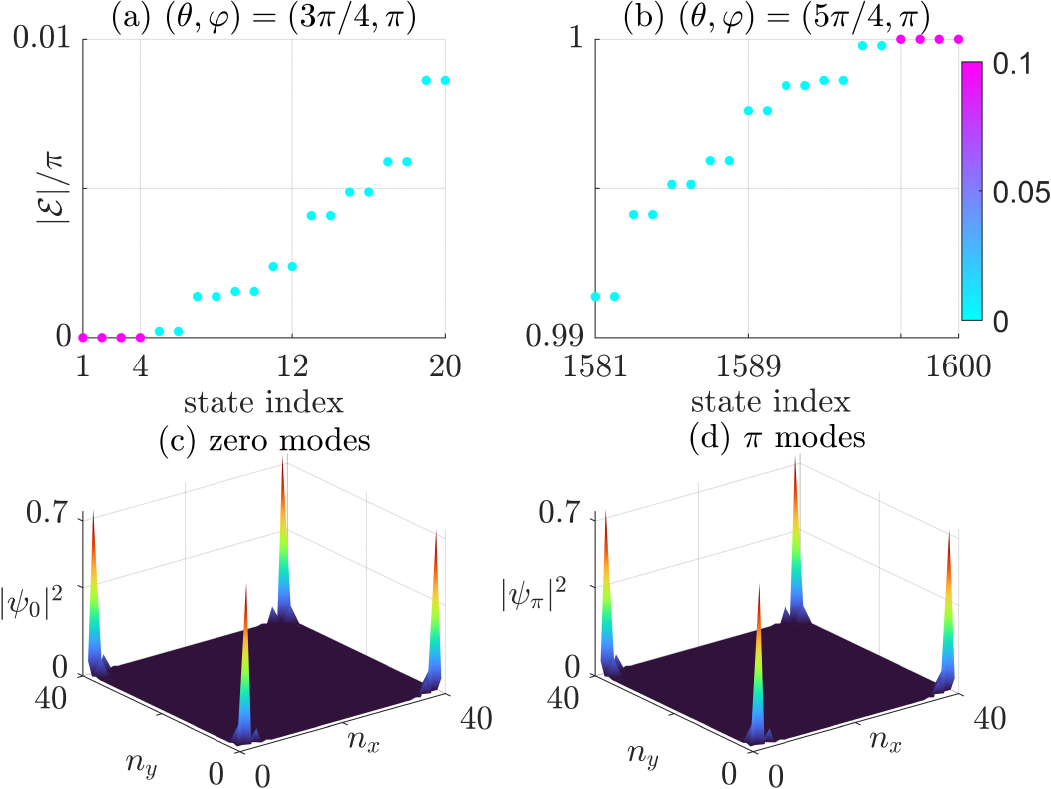}
		\par\end{centering}
	\caption{Floquet corner modes of the kicked CL-SSH model and their probability
		distributions under OBCs. The variables $(\theta,\varphi)$ and related
		system parameters $(J_{x0},J_{x1},J_{y0},J_{y1})$ are defined following
		those of Fig.~\ref{fig:Sketch}(d). (a) shows the spectrum of the
		first twenty eigenstates whose quasienergies are closest to $|{\cal E}|=0$
		at $(\theta,\varphi)=(3\pi/4,\pi)$. (b) shows the spectrum of the
		first twenty eigenstates whose quasienergies are closest to $|{\cal E}|=\pi$
		at $(\theta,\varphi)=(5\pi/4,\pi)$. (a) and (b) share the same color
		bar, which gives the IPR of each state. (c) shows the probability
		distributions $|\psi_{0}|^{2}$ of the for eigenstates with $|{\cal E}|=0$
		in (a). (d) shows the probability distributions $|\psi_{\pi}|^{2}$
		of the for eigenstates with $|{\cal E}|=\pi$ in (b). In (c) and (d),
		$n_{x}$ and $n_{y}$ denote lattice indices along $x$ and $y$ directions.
		\label{fig:CLSSH2}}
\end{figure}

To further demonstrate the Floquet corner modes at topological critical
points, we present in Fig.~\ref{fig:CLSSH2} the quasienergies and
probability distributions of these corner modes under OBCs. In 
Figs.~\ref{fig:CLSSH2}(a) and \ref{fig:CLSSH2}(b), we find that there
are indeed four degenerate eigenmodes at ${\cal E}=0$ (${\cal E}=\pi$)
in the quasienergy spectrum when the system parameters are taken at
one representative point along the left (right) dashed line of 
Fig.~\ref{fig:Sketch}(d). Moreover, only the modes at zero and $\pi$
quasienergies are strongly localized in space, as highlighted by their
IPRs. In Figs.~\ref{fig:CLSSH2}(c) and \ref{fig:CLSSH2}(d), we observe
that the zero and $\pi$ eigenmodes are all exponentially localized
around the four corners of the lattice, respectively, forming Floquet
topological corner modes coexisting with a gapless quasienergy bulk.
The emergence of these modes at system corners provide explicit experimental
signatures for detecting gapless higher-order topology and phase transitions
along higher-order topological phase boundaries in Floquet
systems. The numbers of Floquet zero and $\pi$ corner modes in 
Figs.~\ref{fig:CLSSH2}(c) and \ref{fig:CLSSH2}(d) are coincident with
the prediction of invariants $(\omega_{0},\omega_{\pi})$ according
to Eq.~(\ref{eq:CLSSHBBC}), verifying again the Floquet bulk-corner
correspondence.

In conclusion, we have shown that Floquet driving could transform
the trivial critical point or gapped phases of static CL-SSH model
into topologically nontrivial critical lines (phase boundaries), carrying
degenerate corner modes at zero or $\pi$ quasienergies that are protected
by chiral symmetry and coexisted with gapless bulk states. We characterized
such second-order topology by a pair of generalized invariants
$(\omega_{0},\omega_{\pi})$ and established their relationship with
the numbers of zero and $\pi$ corner modes, which work equally well
for both gapped and gapless Floquet SOTPs. One remaining question
is whether we could have zero and $\pi$ corner modes to survive together
along a topological phase boundary. We give affirmative answer to
this question by treating our CL-SSH model under a different driving
protocol in the next section.

\section{Topological phase boundaries with coexisting zero and $\pi$ Floquet
	corner modes\label{sec:FSOTP2}}
	
	\begin{figure*}
		\begin{centering}
			\includegraphics[scale=0.6]{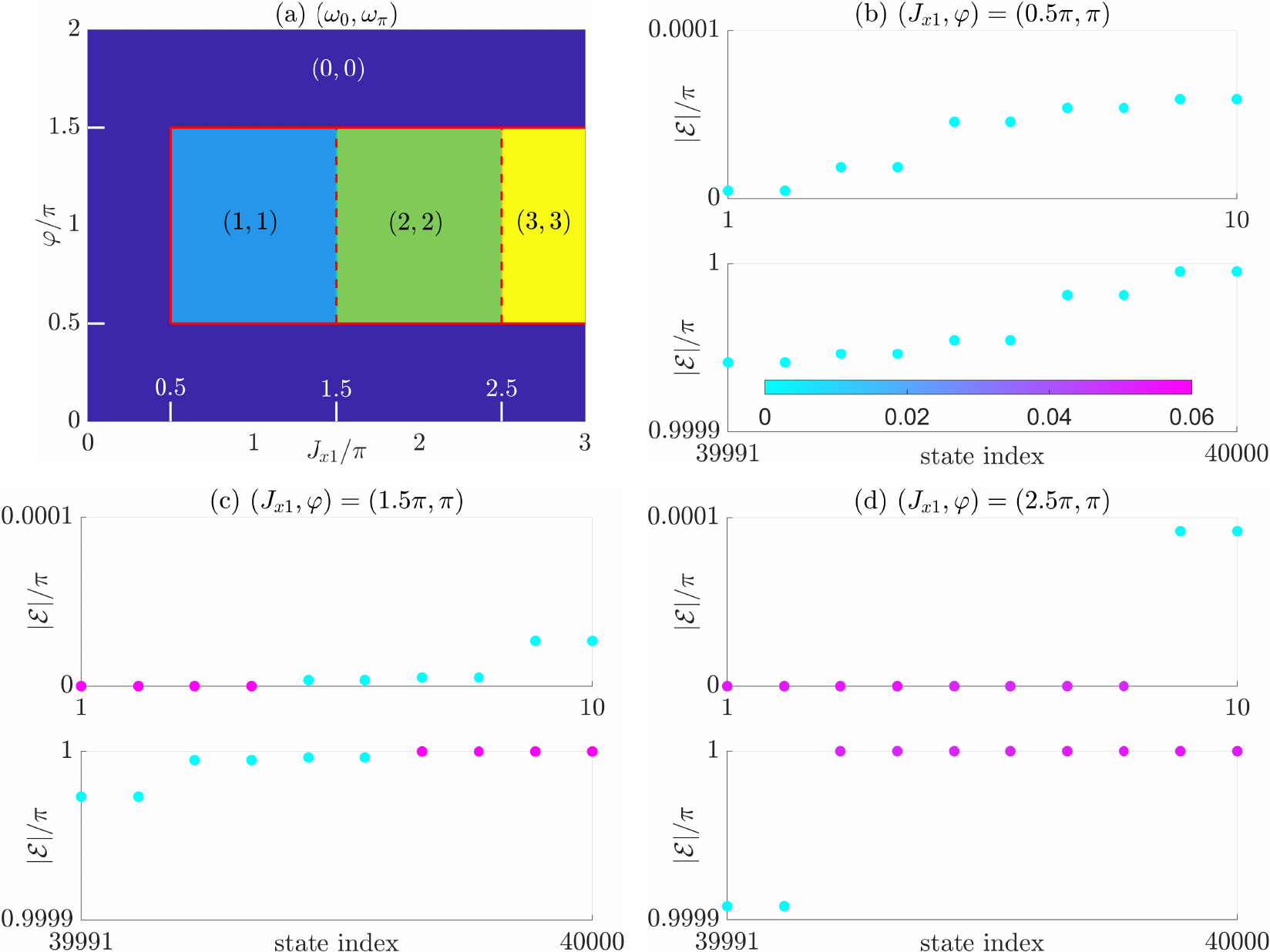}
			\par\end{centering}
		\caption{Topological phase diagram of $U'(k_{x},k_{y})$ {[}Eq.~(\ref{eq:Ukp}){]}
			under PBCs {[}in (a){]} and quasienergy spectra of $U'(k_{x},k_{y})$
			around ${\cal E}=0,\pi$ under OBCs {[}in (b)--(d){]}. Other system
			parameters are $J_{x0}=J'_{x1}=0.5\pi$, $J_{y0}=\frac{\pi}{2}+\frac{\pi}{4}\cos\varphi$
			and $J_{y1}=\frac{\pi}{2}-\frac{\pi}{4}\cos\varphi$ for all panels.
			(b)--(d) share the same color bar, encoding the IPR of each eigenstate.
			The lattice size is set to $L_{x}=L_{y}=200$ for (b)--(d). \label{fig:CLSSHv21}}
	\end{figure*}

In this section, we deal with the possibility of having both zero
and $\pi$ corner modes at the same critical point of Floquet HOTPs.
This is not achievable following the driving protocol of Eqs.~(\ref{eq:Hkt})--(\ref{eq:Uk}).
The key issue is that there are no Floquet phases whose topological
invariants $\omega_{0}$ and $\omega_{\pi}$ are both larger than
one in the periodically kicked CL-SSH model we have studied. To address
this issue, we apply a different driving protocol to each CL in the
CL-SSH model of Fig.~\ref{fig:Sketch}(b), i.e., we take
\begin{equation}
	H'_{x}(k_{x},t)=[J_{x0}+J_{x1}\cos(k_{x})]\sigma_{x}+J'_{x1}\sin(k_{x})\delta_{T}(t)\sigma_{z}.\label{eq:Hxkxtp}
\end{equation}
Compared with the Hamiltonian in Eq.~(\ref{eq:Hxkxt}),
there are two differences. First, the periodic kicks $\delta_{T}(t)\equiv\sum_{\ell\in\mathbb{Z}}\delta(t/T-\ell)$
are now applied to the horizontal NN couplings along each leg of the
CL. Second, the intra-leg NN coupling $J'_{x1}$ and inter-leg second-neighbor
coupling $J_{x1}$ are now allowed to take different values. These
two changes are enough for us to obtain greatly enriched Floquet SOTPs
in the resulting 2D system
\begin{equation}
	H'(k_{x},k_{y},t)=H'_{x}(k_{x},t)\otimes\tau_{0}+\sigma_{0}\otimes H_{y}(k_{y}),\label{eq:Hktp}
\end{equation}
where $H_{y}(k_{y})$ is still given by Eq.~(\ref{eq:Hyky}) of the
SSH model.

Solving the Floquet eigenvalue equation related to $H'_{x}(k_{x},t)$,
we obtain its quasienergy dispersion $\varepsilon(k_{x})=\arccos[\cos(J_{x0}+J_{x1}\cos k_{x})\cos(J'_{x1}\sin k_{x})]$.
Let $\varepsilon(k_{x})=0$ or $\pi$, we find that the quasienergy
gap between Floquet bands $\pm\varepsilon(k_{x})$ vanishes when the
system parameters satisfy the equation
\begin{equation}
	\frac{(\mu\pi-J_{x0})^{2}}{J_{x1}^{2}}+\frac{\nu^{2}\pi^{2}}{(J'_{x1})^{2}}=1,\label{eq:KCLPB2}
\end{equation}
where $\mu,\nu\in\mathbb{Z}$. Specially, the gap closes at quasienergy
zero ($\pi$) if $\mu$ and $\nu$ have the same parity (opposite
parities). It is clear that this phase-boundary equation is more complicated
than Eq.~(\ref{eq:PB}), allowing the Floquet operator of the coupled-ladder
Hamiltonian $H'(k_{x},k_{y},t)$ to yields much richer SOTPs in comparison
with the kicked CL-SSH model of the last section. The underlying physical
reason is that the driving scheme applied in Eq.~(\ref{eq:Hxkxtp})
could induce long-range couplings in the resulting Floquet system,
making it possible for the 2D coupled-ladder to have larger-than-one
topological invariants.

By definition, the Floquet operator of the system described by 
Eq.~(\ref{eq:Hktp}) takes the form
\begin{alignat}{1}
	U'(k_{x},k_{y}) &= e^{-i[J_{x0}+J_{x1}\cos(k_{x})]\sigma_{x}}e^{-iJ'_{x1}\sin(k_{x})\sigma_{z}}\nonumber\\
	&\otimes e^{-iH_{y}(k_{y})},\label{eq:Ukp}
\end{alignat}
where we have taken $\hbar/T=1$ and considered the evolution period
from $t=0^{-}$ to $t=T+0^{-}$. In symmetric time frames, the $U'(k_{x},k_{y})$
in Eq.~(\ref{eq:Ukp}) has the same chiral symmetry as the $U(k_{x},k_{y})$
in Eq.~(\ref{eq:Uk}). Their Floquet SOTPs and bulk-corner correspondence
can thus be characterized by the same theoretical framework according
to Eqs.~(\ref{eq:ome0p}) and (\ref{eq:CLSSHBBC}). In Fig.~\ref{fig:CLSSHv21},
we obtain the topological phase diagram of $U'(k_{x},k_{y})$ under
PBCs and present the Floquet spectra of $U'(k_{x},k_{y})$ around
the center (${\cal E}=0$) and boundary (${\cal E}=\pi$) of the first
quasienergy Brillouin ${\cal E}\in[-\pi,\pi]$ under the OBCs. In
Fig.~\ref{fig:CLSSHv21}(a), we observe one trivial insulating phase
and three Floquet SOTI phases, characterized by the winding numbers
$(\omega_{0},\omega_{\pi})=(0,0)$, $(1,1)$, $(2,2)$ and $(3,3)$
in each uniformly colored region. Notably, these phases are separated
by two distinct types of phase boundaries. The critical line between
trivial and Floquet SOTI phases, highlighted by the red solid line
in Fig.~\ref{fig:CLSSHv21}(a) is itself topologically trivial, featuring
no zero or $\pi$ Floquet corner modes as exemplified in Fig.~\ref{fig:CLSSHv21}(b)
under OBCs. The two critical lines between nontrivial Floquet SOTI
phases, emphasized by the red dashed lines in Fig.~\ref{fig:CLSSHv21}(a)
are yet topologically nontrivial, carrying fourfold-degenerate Floquet
corner modes at both zero and $\pi$ quasienergies as showcased in
Figs.~\ref{fig:CLSSHv21}(c) and \ref{fig:CLSSHv21}(d) under OBCs.
There are four (eight) zero modes and four (eight) $\pi$ modes in
Fig.~\ref{fig:CLSSHv21}(c) (Fig.~\ref{fig:CLSSHv21}(d)), whose IPRs
are clearly larger than other states close to their quasienergies.
Moreover, the numbers of zero and $\pi$ Floquet corner modes at the
critical points of $U'(k_{x},k_{y})$ are correctly counted by the
invariants $(\omega_{0},\omega_{\pi})$ according to Eq.~(\ref{eq:CLSSHBBC}),
which verifies the rule of bulk-corner correspondence we proposed
for 2D, chiral-symmetric driven systems.

\begin{figure}
	\begin{centering}
		\includegraphics[scale=0.51]{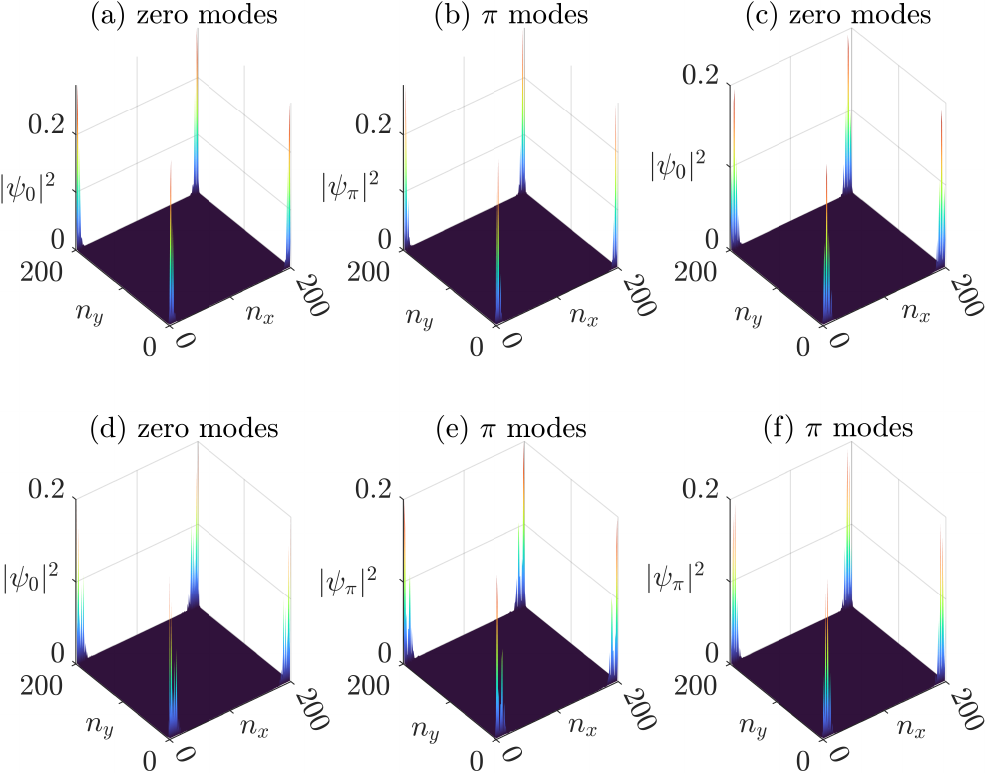}
		\par\end{centering}
	\caption{Probability distributions of the corner modes of $U'(k_{x},k_{y})$
		at critical points under OBCs. (a) shows the four Floquet states at
		${\cal E}=0$ in Fig.~\ref{fig:CLSSHv21}(c). (b) shows the four Floquet
		states at ${\cal E}=\pi$ in Fig.~\ref{fig:CLSSHv21}(c). (c) and
		(d) show the eight Floquet states at ${\cal E}=0$ in Fig.~\ref{fig:CLSSHv21}(d).
		(e) and (f) show the eight Floquet states at ${\cal E}=\pi$ in 
		Fig.~\ref{fig:CLSSHv21}(d). \label{fig:CLSSHv22}}
\end{figure}

To further illustrate the spatial configurations of Floquet corner
modes at topological phase boundaries, we show in Figs.~\ref{fig:CLSSHv22}(a)--\ref{fig:CLSSHv22}(b)
and Figs.~\ref{fig:CLSSHv22}(c)--\ref{fig:CLSSHv22}(f) the probability
distributions of the zero and $\pi$ eigenmodes in Figs.~\ref{fig:CLSSHv21}(c)
and \ref{fig:CLSSHv21}(d), respectively. We find that these zero
and $\pi$ modes are indeed corner-localized, irrespective of the
bulk-gap closing at zero and $\pi$ quasienergies. These corner states
could thus survive topological transitions between different Floquet
SOTI phases, offering characteristic signatures of gapless higher-order
topology at Floquet critical points. Their coexistence at zero and
$\pi$ quasienergies further implies that they have no static counterparts
and are unique to nonequilibrium Floquet settings. In principle, we
could obtain as many coexisting zero and $\pi$ corner modes as we
want at the critical points of the system described by $U'(k_{x},k_{y})$
via tuning its parameters, i.e., $J_{x1}$. Therefore, Floquet driving
offers a flexible tool for us to enrich the higher-order topology
at topological transition points and induce higher-order gSPT phases
that are intrinsically nonequilibrium.

\section{Summary and Discussion\label{sec:Sum}}

In this work, we revealed the nontrivial topology and corner modes
at the critical points between different Floquet SOTPs. Focusing on
2D systems with chiral symmetry, we uncovered the conditions for degenerate
Floquet corner modes to appear at zero and $\pi$ quasienergies when
the bulk states remain gapless. Moreover, we analytically characterized
the transitions between different Floquet SOTPs with gapless bulk
spectra. To resolve the issue of bulk-corner correspondence, we proposed
a pair of generalized winding numbers under the PBC based on the Cauchy's
argument principle, which could correctly count the numbers of zero
and $\pi$ corner modes in both gapped and gapless Floquet SOTPs so
long as the chiral symmetry is preserved. Going beyond the minimal
model, we demonstrated the possibility of having as many zero and
$\pi$ corner modes as possible to coexist with a fully gapless quasienergy
bulk, which is enabled by long-range couplings induced via Floquet
driving. Our work not only extended the study of higher-order topology
and bulk-corner correspondence to the critical points of nonequilibrium
systems, but also unveiled the richness of Floquet gapless topology
beyond first-order topological phases.

For our theory of gapless higher-order topology to work, the following conditions are necessary. In the static case, we need both the $H_x$ and $H_y$ in Eq.~(\ref{eq:H}) to be chiral symmetric, two-band Bloch Hamiltonians. Their chiral symmetries can have different representations. Under these constraints, the definition of topological invariant $\omega=\omega_x \omega_y$ in Eq.~(\ref{eq:ome}) and the bulk-corner correspondence $N_0=4\omega$ in Eq.~(\ref{eq:BBC}) are generic and independent of the specific forms of subsystem Hamiltonians. In the Floquet case, we need both the Floquet operator $U_x$ generated by the $H_x(t)$ in Eq.~(\ref{eq:Hkt}) and the Hamiltonian $H_y$ in Eq.~(\ref{eq:Hkt}) to be two-band models with chiral symmetries. Under these constraints, the definition of topological invariants $(\omega_0,\omega_\pi)$ in Eq.~(\ref{eq:ome0p}) and the rule of bulk-corner correspondence $(N_0,N_\pi)=4(\omega_0,\omega_\pi)$ in Eq.~(\ref{eq:CLSSHBBC}) are generic and independent of the specific forms of subsystem Floquet operators. The models we treated in Secs.~\ref{sec:SOTP} and \ref{sec:FSOTP} thus correspond to the ``theoretical minimum'' of static and Floquet SOTPs with topologically trivial and nontrivial critical points, respectively.

By construction, both our static and Floquet models have the rectangular lattice geometry. This geometry enforces the corner modes of our systems to appear at four corners of the lattice. Despite this built-in geometric constraint, the chiral symmetries are the only indispensable symmetries for the protection of topological corner modes in our systems. Our definition of topological invariants $(\omega_0,\omega_{\pi})$ requires the Floquet operator $U(k_x,k_y)$ to have the factorization structure $U(k_x)\otimes U(k_y)$. Besides the chiral symmetries of $U(k_x)$ and $U(k_y)$, other internal (e.g., time-reversal, particle-hole) and lattice (such as reflection) symmetries are not essential for protecting Floquet corner modes, making our approach flexible in realizing gapless higher-order Floquet topology in more complicated situations. Our scheme can also be generalized to construct $n$th-order Floquet topological phases (either gapped or gapless) in $n$-dimensions for any $n>2$. For example, the Floquet model $U(k_x,k_y,k_z)=U(k_x)\otimes U(k_y)\otimes U(k_z)$ could possess zero and $\pi$ corner modes if the $U(k_i)$ has chiral symmetry for all $i=x,y,z$. When the tensor product structure in $U(k_x,k_y)=U(k_x)\otimes U(k_y)$ is lifted, the topological invariants we introduced cannot be directly computed following our scheme, even though their associated corner modes show clear robustness to perturbations that break both the tensor product structure and some other lattice symmetries, as we illustrated in the Appendix E. A next step in the study of gapless Floquet topology is thus to extend the definition of topological invariants we introduced to other higher-order Floquet models that do not rely on the factorization structure of Floquet operator in their constructions.

In future studies, it would be important to systematically check the
stability of higher-order Floquet topology at critical points against
disorder, lattice quasiperiodicity \cite{RoyPRB2022} and many-body interactions. Floquet gapless higher-order
topology beyond two spatial dimensions and in other symmetry classes
are also interesting research topics. Finally, the experimental detection
of Floquet higher-order topological criticality in quantum simulators
like cold atoms, photonic and acoustic systems merits further explorations.

\begin{acknowledgments}
	This work is supported by the National Natural Science Foundation of China (Grants No.~12275260 and No.~11905211), the Fundamental Research Funds for the Central Universities (Grant No.~202364008), and the Young Talents Project of Ocean University of China.
\end{acknowledgments}
\vspace{0.5cm}

\appendix

\section{Edge and corner zero modes of static models}\label{sec:App1}

In this Appendix, we compute exact solutions of zero-energy eigenmodes
in the 1D CL and SSH models. We show that they can be exponentially
localized around system edges only in the gapped topological phases.
These zero-mode solutions will be further used to construct corner
zero modes of the coupled 2D CL-SSH model, as discussed in the main
text. Explicit conditions for these eigenmodes to be localized exponentially
at corners will be given.

After Fourier transformation, the Bloch Hamiltonian $H_{x}(k_{x})$
of CL in the main text can be expressed in real space as
\begin{alignat}{1}
	H_{x}&=  J_{x0}\sum_{m}(\hat{u}_{m}^{\dagger}\hat{v}_{m}+{\rm H.c.})\nonumber \\
	&+  \frac{J_{x1}}{2}\sum_{m}[(\hat{u}_{m}^{\dagger}\hat{v}_{m+1}+\hat{u}_{m+1}^{\dagger}\hat{v}_{m})+{\rm H.c.}]\nonumber \\
	&+  \frac{J_{x1}}{2i}\sum_{m}[(\hat{u}_{m}^{\dagger}\hat{u}_{m+1}-\hat{v}_{m}^{\dagger}\hat{v}_{m+1})-{\rm H.c.}],\label{eq:Hx}
\end{alignat}
where $m$ is the unit-cell index along $x$-direction. $\hat{u}_{m}^{\dagger}$
and $\hat{v}_{m}^{\dagger}$ create a fermion in the sublattices $u$
and $v$ of the $m$th unit cell on the upper and lower legs of the
ladder {[}see Fig.~\ref{fig:Sketch}(a){]}, respectively. Consider
a half-infinite ladder with $m=1,2,...,\infty$. A general solution
of the eigenvalue equation $H_{x}|\psi\rangle=E|\psi\rangle$ can
be expanded as $|\psi\rangle=\sum_{m}(\psi_{m,u}\hat{u}_{m}^{\dagger}+\psi_{m,v}\hat{v}_{m}^{\dagger})|\emptyset\rangle$,
where $|\emptyset\rangle$ denotes the vacuum state. If $|\psi\rangle$
is a zero-energy eigenstate of $H_{x}$, it must satisfy $H_{x}|\psi\rangle=0$,
yielding
\begin{widetext}
	\begin{alignat}{1}
		& J_{x0}\sum_{m}(\psi_{m,v}\hat{u}_{m}^{\dagger}+\psi_{m,u}\hat{v}_{m}^{\dagger})
		+ \frac{J_{x1}}{2}\sum_{m}(\psi_{m+1,v}\hat{u}_{m}^{\dagger}+\psi_{m,v}\hat{u}_{m+1}^{\dagger}+\psi_{m+1,u}\hat{v}_{m}^{\dagger}+\psi_{m,u}\hat{v}_{m+1}^{\dagger})\nonumber \\
		&+  \frac{J_{x1}}{2i}\sum_{m}(\psi_{m+1,u}\hat{u}_{m}^{\dagger}-\psi_{m,u}\hat{u}_{m+1}^{\dagger}-\psi_{m+1,v}\hat{v}_{m}^{\dagger}+\psi_{m,v}\hat{v}_{m+1}^{\dagger})=0.\label{eq:Ex}
	\end{alignat}
\end{widetext}
Solving the difference equation for the coefficients $\psi_{m,u}$
and $\psi_{m,v}$, we find for any $m\in\mathbb{Z}^{+}$ that
\begin{alignat}{1}
	\psi_{m,u}&=i\psi_{m,v},\qquad\psi_{m,u}=\left(-\frac{J_{x0}}{J_{x1}}\right)^{m-1}\psi_{1,u},\nonumber\\
	\psi_{m,v}&=\left(-\frac{J_{x0}}{J_{x1}}\right)^{m-1}\psi_{1,v}.\label{eq:cx}
\end{alignat}
Therefore, up to a normalization constant, the wavefunction of zero-energy
solution takes the form
\begin{equation}
	|\psi_{{\rm L}}\rangle=\sum_{m=1}^{\infty}\left(-\frac{J_{x0}}{J_{x1}}\right)^{m-1}(\hat{u}_{m}^{\dagger}-i\hat{v}_{m}^{\dagger})|\emptyset\rangle,\label{eq:0x1}
\end{equation}
where ``L'' means left. It is clear that $|\psi_{{\rm L}}\rangle$
represents an edge zero mode localized exponentially from site $m=1$
if and only if $|J_{x1}|>|J_{x0}|$. At the critical point $|J_{x1}|=|J_{x0}|$,
$|\psi_{{\rm L}}\rangle$ becomes extended uniformly across all sites
of the two legs. Another edge zero mode $|\psi_{{\rm R}}\rangle$
(``R'' means right) can be found by considering the half-infinite
chain with unit-cell indices $m=-\infty,...,M-1,M$ ($M\gg1$) and
solving $H_{x}|\psi_{{\rm R}}\rangle=0$, yielding
\begin{equation}
	|\psi_{{\rm R}}\rangle=\sum_{m=-\infty}^{M}\left(-\frac{J_{x0}}{J_{x1}}\right)^{M-m}(\hat{u}_{m}^{\dagger}+i\hat{v}_{m}^{\dagger})|\emptyset\rangle.\label{eq:0x2}
\end{equation}
The two modes $|\psi_{{\rm L}}\rangle$ and $|\psi_{{\rm R}}\rangle$
satisfy $\langle\psi_{{\rm L}}|\psi_{{\rm R}}\rangle=0$, making them
degenerate at $E=0$.

Upon Fourier transformation, the Bloch Hamiltonian $H_{y}(k_{y})$
of SSH model can be expressed in real space as
\begin{equation}
	H_{y}=\sum_{n}\left(J_{y0}\hat{a}_{n}^{\dagger}\hat{b}_{n}+J_{y1}\hat{b}_{n}^{\dagger}\hat{a}_{n+1}+{\rm H.c.}\right),\label{eq:Hy}
\end{equation}
where $n$ is the unit-cell index along $y$-direction. $\hat{a}_{n}^{\dagger}$
and $\hat{b}_{n}^{\dagger}$ creates a fermion on the sublattice $a$
and $b$ of the $n$th unit cell, respectively {[}see Fig.~\ref{fig:Sketch}(a){]}.
Consider a half-infinite chain with $n=1,2,...,\infty$, a general
solution of the eigenvalue equation $H_{y}|\psi'\rangle=E|\psi'\rangle$
can be expanded as $|\psi'\rangle=\sum_{n}(\varphi_{n,a}\hat{a}_{n}^{\dagger}+\varphi_{n,b}\hat{b}_{n}^{\dagger})|\emptyset\rangle$,
where $|\emptyset\rangle$ denotes the vacuum state. If $|\psi'\rangle$
is a zero-energy eigenstate of $H_{y}$, it must satisfy $H_{y}|\psi'\rangle=0$,
yielding
\begin{widetext}
	\begin{equation}
		\sum_{n}\left(J_{y0}\varphi_{n,b}\hat{a}_{n}^{\dagger}+J_{y0}\varphi_{n,a}\hat{b}_{n}^{\dagger}+J_{y1}\varphi_{n+1,a}\hat{b}_{n}^{\dagger}+J_{y1}\varphi_{n,b}\hat{a}_{n+1}^{\dagger}\right)=0.\label{eq:Ey}
	\end{equation}
\end{widetext}
Solving the difference equation for the coefficients $\varphi_{n,a}$
and $\varphi_{n,b}$, we find for any $n\in\mathbb{Z}^{+}$ that
\begin{equation}
	\varphi_{n,a}=\left(-\frac{J_{y0}}{J_{y1}}\right)^{n-1}\varphi_{1,a},\qquad\varphi_{n,b}=0.\label{eq:cy}
\end{equation}
Therefore, up to a normalization constant, the wavefunction of zero-energy
solution takes the form
\begin{equation}
	|\psi'_{{\rm B}}\rangle=\sum_{n=1}^{\infty}\left(-\frac{J_{y0}}{J_{y1}}\right)^{n-1}\hat{a}_{n}^{\dagger}|\emptyset\rangle,\label{eq:0y1}
\end{equation}
where ``B'' denotes bottom. It is clear that $|\psi'_{{\rm B}}\rangle$
represents an edge zero mode localized exponentially along $y$-direction
from the site $n=1$ if and only if $|J_{y1}|>|J_{y0}|$. At the critical
point $|J_{y1}|=|J_{y0}|$, $|\psi'_{{\rm B}}\rangle$ becomes a charge
density wave, which is extended uniformly across all the $a$ sublattices
of the system. Another edge zero mode $|\psi'_{{\rm T}}\rangle$ (``T''
denotes top) can be found by considering the half-infinite chain with
unit-cell indices $n=-\infty,...,N-1,N$ ($N\gg1$) and solving $H_{y}|\psi'\rangle=0$,
yielding
\begin{equation}
	|\psi'_{{\rm T}}\rangle=\sum_{n=-\infty}^{N}\left(-\frac{J_{y0}}{J_{y1}}\right)^{N-n}\hat{b}_{n}^{\dagger}|\emptyset\rangle.\label{eq:0y2}
\end{equation}
The two modes $|\psi'_{{\rm B}}\rangle$ and $|\psi'_{{\rm T}}\rangle$
satisfy $\langle\psi'_{{\rm B}}|\psi'_{{\rm T}}\rangle=0$, affirming
their degeneracy at $E=0$. The conditions for them to be localized
edge states are the same.

In two spatial dimensions, we can now construct four zero modes from
the zero modes of 1D CL and SSH models. Their wave functions (up to
normalization factors) are given by
\begin{alignat}{1}
& |\Psi_{{\rm LB}}\rangle=|\psi_{{\rm L}}\rangle\otimes|\psi'_{{\rm B}}\rangle,\quad|\Psi_{{\rm LT}}\rangle=|\psi_{{\rm L}}\rangle\otimes|\psi'_{{\rm T}}\rangle,\label{eq:C0M1}\\
& |\Psi_{{\rm RB}}\rangle=|\psi_{{\rm R}}\rangle\otimes|\psi'_{{\rm B}}\rangle,\quad|\Psi_{{\rm RT}}\rangle=|\psi_{{\rm R}}\rangle\otimes|\psi'_{{\rm T}}\rangle.\label{eq:C0M2}
\end{alignat}
When the conditions $|J_{x1}|>|J_{x0}|$ and $|J_{y1}|>|J_{y0}|$
are both satisfied, these four zero modes are exponentially localized
at the left-bottom (LB), left-top (LT), right-bottom (RB) and right-top
(RT) corners of the 2D lattice, respectively. These corner modes will
vanish in other parameter domains, including the critical regions
where the bulk spectra are gapless. Their fourfold degeneracy is
protected by the chiral symmetry of the 2D system.

\section{Topology and bulk-edge correspondence of the kicked CL}\label{sec:App2}

\begin{figure}
	\begin{centering}
		\includegraphics[scale=0.485]{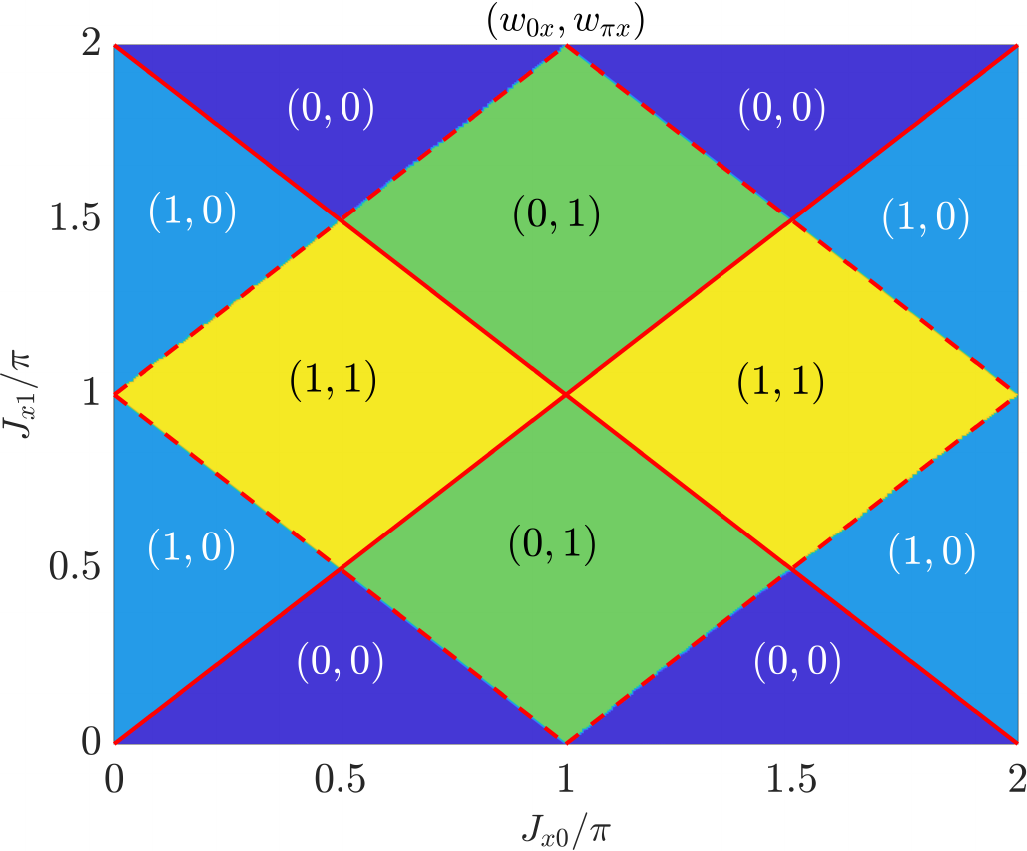}
		\par\end{centering}
	\caption{Topological phase diagram of the 1D kicked CL. Each region with a
		uniform color corresponds to a gapped Floquet phase, with the winding
		numbers $(w_{0x},w_{\pi x})$ denoted therein. The solid and dashed
		lines are given by Eq.~(\ref{eq:PB}), along which the bulk quasienergy
		gaps of $U_{x}(k_{x})$ close at $E=0$ and $E=\pm\pi$, respectively.
		\label{fig:KCLPD}}
\end{figure}

In this Appendix, we establish the bulk-edge correspondence for the
kicked CL from two complementary perspectives. Focusing on the edge,
we obtain exact solutions at zero and $\pi$ quasienergies for a half-infinite
ladder, and reveal under what condition they could become exponentially
localized edge modes. Focusing on the bulk, we generalize the definition
of winding numbers following the strategy leading to Eq.~(\ref{eq:omex})
\cite{FgSPT01}, and verify that our generalized winding numbers could
correctly count the zero and $\pi$ Floquet edge modes in both the
gapped phases and at the critical points, thereby establishing bulk-edge
correspondence throughout the parameter space of the system. These
two perspectives yield consistent results.

As a start, we obtain the topological phase diagram of $U_{x}(k_{x})$
{[}Eq.~(\ref{eq:Uxkx}){]} for its gapped phases by evaluating the
more conventional winding numbers in Eq.~(\ref{eq:w0px}). We can
expand the Floquet operator $U_{\alpha x}(k_{x})$ {[}Eqs.~(\ref{eq:U1xkx})
and (\ref{eq:U2xkx}){]} in symmetric time frame $\alpha$ as
\begin{equation}
	U_{\alpha x}(k_{x})=\cos[\varepsilon(k_{x})]-i(d_{x\alpha}\sigma_{x}+d_{z\alpha}\sigma_{z}),\,\,\alpha=1,2.\label{eq:UaExp}
\end{equation}
The expression of $\varepsilon(k_{x})$ is given by Eq.~(\ref{eq:Ekx}).
The components in front of Pauli matrices $\sigma_{x}$ and $\sigma_{z}$
are
\begin{widetext}
	\begin{alignat}{1}
		d_{x1}&=\sin J_{x0}\cos^{2}\frac{J_{x1}}{2}\cos(\lambda'k_{x})+\cos J_{x0}\sin J_{x1}\cos(\lambda k_{x})-\sin J_{x0}\sin^{2}\frac{J_{x1}}{2}\cos[(2\lambda-\lambda')k_{x}],\label{eq:dx1}\\
		d_{z1}&=\sin J_{x0}\cos^{2}\frac{J_{x1}}{2}\sin(\lambda'k_{x})+\cos J_{x0}\sin J_{x1}\sin(\lambda k_{x})-\sin J_{x0}\sin^{2}\frac{J_{x1}}{2}\sin[(2\lambda-\lambda')k_{x}],\label{eq:dz1}\\
		d_{x2}&=\sin J_{x1}\cos^{2}\frac{J_{x0}}{2}\cos(\lambda k_{x})+\cos J_{x1}\sin J_{x0}\cos(\lambda'k_{x})-\sin J_{x1}\sin^{2}\frac{J_{x0}}{2}\cos[(2\lambda'-\lambda)k_{x}],\label{eq:dx2}\\
		d_{z2}&=\sin J_{x1}\cos^{2}\frac{J_{x0}}{2}\sin(\lambda k_{x})+\cos J_{x1}\sin J_{x0}\sin(\lambda'k_{x})-\sin J_{x1}\sin^{2}\frac{J_{x0}}{2}\sin[(2\lambda'-\lambda)k_{x}],\label{eq:dz2}
	\end{alignat}
\end{widetext}
where $\lambda=1$ and $\lambda'=0$. Plugging these expressions into
Eq.~(\ref{eq:w12x}) gives us the winding numbers $(w_{1x},w_{2x})$,
whose combinations further yield the topological invariants $(w_{0x},w_{\pi x})$
of $U_{x}(k_{x})$ through Eq.~(\ref{eq:w0px}). In Fig.~\ref{fig:KCLPD},
we report the numerical results of $(w_{0x},w_{\pi x})$ in the parameter
space $(J_{x0},J_{x1})\in[0,2\pi]\times[0,2\pi]$ of the kicked CL.
As the $U_{x}(k_{x})$ is invariant under the change of $J_{x0}$
or $J_{x1}$ by $2\pi$, the phase diagrams in other parameter regions
can be obtained by translating Fig.~\ref{fig:KCLPD} along the $J_{x0}$
and $J_{x1}$ axes by integer multiples of $2\pi$. Moreover, we notice
that the lower-left quarter of Fig.~\ref{fig:KCLPD}, with $(J_{x0},J_{x1})\in[0,\pi]\times[0,\pi]$,
already contains all possible topological phases of the system. Other
parts of Fig.~\ref{fig:KCLPD} can be obtained
by reflecting this lower-left quarter along the parameter axes $J_{x0}=\pi$
and $J_{x1}=\pi$. We thus focus on characterizing the topological
phases and transitions in the quarter of Fig.~\ref{fig:KCLPD} with
$(J_{x0},J_{x1})\in[0,\pi]\times[0,\pi]$ in the remaining part
of this Appendix. General conclusions for all parameter regions
of the kicked CL will be summarized in Appendix~\ref{sec:App4}.

Next, we propose a generalization of the topological winding numbers
$(w_{0x},w_{\pi x})$ based on the Cauchy's argument principle \cite{gSPT02},
which would be applicable in both the gapped and gapless
regions \cite{FgSPT01}. In Eq.~(\ref{eq:UaExp}), we identify
the Floquet effective Hamiltonian in the symmetric time frame $\alpha$
as $H_{\alpha x}(k_{x})=[U_{\alpha x}^{\dagger}(k_{x})-U_{\alpha x}(k_{x})]/(2i)$,
i.e.,
\begin{equation}
	H_{\alpha x}(k_{x})=d_{x\alpha}\sigma_{x}+d_{z\alpha}\sigma_{z},\quad\alpha=1,2.\label{eq:Heff}
\end{equation}
As $[H_{\alpha x}(k_{x}),U_{\alpha x}(k_{x})]=0$, the
$H_{\alpha x}(k_{x})$ shares the same eigenstates and symmetries
with the original Floquet operator $U_{\alpha x}(k_{x})$. Meanwhile,
when the Floquet bands of $U_{\alpha x}(k_{x})$ touch at the quasienergy
zero or $\pi$, the spectrum of $H_{\alpha x}(k_{x})$ becomes gapless
at $\varepsilon=0$. Therefore, the critical points of $U_{\alpha x}(k_{x})$
are the same as its effective Hamiltonian $H_{\alpha x}(k_{x})$ in
the $J_{x0}$-$J_{x1}$ parameter space, even though they appear at
different (quasi)energies.

With Eqs.~(\ref{eq:UaExp})--(\ref{eq:dz2}) and following Ref.~\cite{gSPT02},
we find the complex mapping functions of effective Hamiltonians $H_{1x}(k_{x})$
and $H_{2x}(k_{x})$ as
\begin{widetext}
	\begin{alignat}{1}
		f_{1x}(z)= & \sin J_{x0}\cos^{2}\frac{J_{x1}}{2}+\cos J_{x0}\sin J_{x1}z-\sin J_{x0}\sin^{2}\frac{J_{x1}}{2}z^{2},\nonumber \\
		f_{2x}(z)= & \frac{\left(\sin J_{x1}\cos^{2}\frac{J_{x0}}{2}z^{2}+\cos J_{x1}\sin J_{x0}z-\sin J_{x1}\sin^{2}\frac{J_{x0}}{2}\right)z}{z^{2}}.\label{eq:f12xz}
	\end{alignat}
\end{widetext}
The Cauchy's argument principle then states that for
$f_{\alpha x}(z)$, the difference between the number of its zeros
($N_{\alpha z}$) and poles ($N_{\alpha p}$) (including their multiplicities
and orders) inside the unit circle $|z|=1$ is an integer-quantized
winding number, which will be denoted by $\omega_{\alpha x}$ ($\alpha=1,2$).
However, this zero-pole counting rule needs to be modified
along the critical lines where the gaps close at quasienergy $\pi$
\cite{FgSPT01}. Using $(\omega_{1x},\omega_{2x})$, we finally
arrive at a pair of generalized topological invariants as
\begin{equation}
	(\omega_{0x},\omega_{\pi x})=\frac{1}{2}(\omega_{1x}+\omega_{2x},\omega_{1x}-\omega_{2x}).\label{eq:ome0px}
\end{equation}

We now verify that the $(\omega_{0x},\omega_{\pi x})$
in Eq.~(\ref{eq:ome0px}) could reproduce the topological phase diagram
in Fig.~\ref{fig:KCLPD}, i.e., we have $(\omega_{0x},\omega_{\pi x})=(w_{0x},w_{\pi x})$
in all the gapped phases. Since the winding numbers could not change
when the spectrum gap remains open, we can just do the zero-pole counting
at one representative point within each gapped parameter region in
order to characterize the corresponding phase. Without losing
generality, we consider the representative points $(J_{x0},J_{x1})=(\pi/2,0)$,
$(0,\pi/2)$, $(\pi,\pi/2)$ and $(\pi/2,\pi)$ on the lower-left
quarter of Fig.~\ref{fig:KCLPD}. The zero-pole counting for the $f_{1x}(z)$
and $f_{2x}(z)$ in Eq.~(\ref{eq:f12xz}) can then be performed directly.
For example, at $(J_{x0},J_{x1})=(\pi/2,\pi)$, we have $f_{1x}(z)=z^{2}$
and $f_{2x}(z)=-1$, yielding $N_{1z}-N_{1p}=\omega_{1x}=2$ and $N_{2z}-N_{2p}=\omega_{2x}=0$.
From Eq.~(\ref{eq:ome0px}), we then obtain $(\omega_{0x},\omega_{\pi x})=(1,1)$,
which are exactly identical to the winding numbers $(w_{0x},w_{\pi x})$
shown in Fig.~\ref{fig:KCLPD}. One can repeat the calculations at
other representative points. The final results confirm that for all
the gapped phases, we have $(\omega_{0x},\omega_{\pi x})=(w_{0x},w_{\pi x})$.
The generalized winding numbers $(\omega_{0x},\omega_{\pi x})$ thus
predict the same phase diagram as $(w_{0x},w_{\pi x})$ for gapped
topological phases of the kicked CL.

Along the critical lines, the winding numbers $(w_{0x},w_{\pi x})$
can take half-integer values and may become ill-defined.
Due to the Cauchy's argument principle, the generalized
invariants $(\omega_{0x},\omega_{\pi x})$ are still
integers. To see this, we first consider the critical line $J_{x1}=J_{x0}$
within the lower-left quarter of Fig.~\ref{fig:KCLPD} {[}i.e., with
$J_{x0}\in(0,\pi)${]}. There is a multi-critical point $J_{x1}=J_{x0}=\pi/2$
along this critical line, which divides it into two segments. Within
the segment $J_{x0}\in(0,\pi/2)$ or $J_{x0}\in(\pi/2,\pi)$, the
gap structure of the system remains unchanged, and we can pick up
one representative point in each segment to analyze the generalized
winding numbers $(\omega_{0x},\omega_{\pi x})$. At $(J_{x0},J_{x1})=(\pi/3,\pi/3)$,
we find $\omega_{1x}\equiv N_{1z}-N_{1p}=0$ and $\omega_{2x}\equiv N_{2z}-N_{2p}=0$,
yielding $(\omega_{0x},\omega_{\pi x})=(0,0)$ due to Eq.~(\ref{eq:ome0px}),
which is valid along $J_{x1}=J_{x0}$ for $J_{x0}\in(0,\pi/2)$.
At $(J_{x0},J_{x1})=(2\pi/3,2\pi/3)$, we find $\omega_{1x}\equiv N_{1z}-N_{1p}=1$
and $\omega_{2x}\equiv N_{2z}-N_{2p}=-1$, yielding $(\omega_{0x},\omega_{\pi x})=(0,1)$
due to Eq.~(\ref{eq:ome0px}), which is valid along $J_{x1}=J_{x0}$
for $J_{x0}\in(\pi/2,\pi)$. Notably, we realize that the two segments
$J_{x0}\in(0,\pi/2)$ and $J_{x0}\in(\pi/2,\pi)$ along the the critical
line $J_{x1}=J_{x0}$ are topologically discriminated by the invariant
$\omega_{\pi x}$. Next, we consider the critical line
$J_{x1}=\pi-J_{x0}$ within the lower-left quarter of Fig.~\ref{fig:KCLPD},
along which the quasienergy gap closes at $\pm\pi$ and a refined
zero-pole counting rule is needed \cite{FgSPT01}. Specially, at the
representative point $(J_{x0},J_{x1})=(\pi/3,2\pi/3)$, we find $\omega_{1x}\equiv N_{1z}-N_{1p}/2=1$
and $\omega_{2x}\equiv N_{2z}-N_{2p}/2=1$, yielding $(\omega_{0x},\omega_{\pi x})=(1,0)$
due to Eq.~(\ref{eq:ome0px}), which is valid along $J_{x1}=\pi-J_{x0}$
for $J_{x0}\in(0,\pi/2)$. At another representative point $(J_{x0},J_{x1})=(2\pi/3,\pi/3)$,
we find $\omega_{1x}\equiv N_{1z}-N_{1p}/2=0$ and $\omega_{2x}\equiv N_{2z}-N_{2p}/2=0$,
yielding $(\omega_{0x},\omega_{\pi x})=(0,0)$ due to Eq.~(\ref{eq:ome0px}),
which is valid along $J_{x1}=\pi-J_{x0}$ for $J_{x0}\in(\pi/2,\pi)$.
We see that the two segments $J_{x0}\in(0,\pi/2)$ and $J_{x0}\in(\pi/2,\pi)$
along the the critical line $J_{x1}=\pi-J_{x0}$ are also topologically
inequivalent, with the generalized invariant $\omega_{0x}$ taking
different values. Finally, at the multi-critical point $(J_{x0},J_{x1})=(\pi/2,\pi/2)$,
the spectral gaps close at both zero and $\pi$ quasienergies. We
find $\omega_{1x}\equiv N_{1z}-N_{1p}/2=0$ and $\omega_{2x}\equiv N_{2z}-N_{2p}/2=0$,
yielding $(\omega_{0x},\omega_{\pi x})=(0,0)$. 

To sum up, we propose the following definitions for the generalized
winding numbers $(\omega_{1x},\omega_{2x})$ in two symmetric time
frames
\begin{equation}
	\omega_{\alpha x}=\begin{cases}
		N_{\alpha z}-N_{\alpha p}, & \Delta_{\pi}\neq0,\\
		N_{\alpha z}-N_{\alpha p}/2, & \Delta_{\pi}=0,
	\end{cases}\quad\alpha=1,2,\label{eq:ome12x}
\end{equation}
where $\Delta_{\pi}$ measures the spectral gap of $U_{x}(k_{x})$
at quasienergy $\pi$. Following this definition, the invariants $(\omega_{0x},\omega_{\pi x})$
in Eq.~(\ref{eq:ome0px}) are always integer quantized, as demonstrated
in the above discussions. Moreover, they could completely characterize
the bulk-edge correspondence of the kicked CL. This point will be
confirmed below from the perspective of edge states.

To simplify the calculation of edge states, we first make a basis
transformation on the Floquet operator $U_{x}=\sum_{k_{x}}\hat{\Psi}_{k_{x}}^{\dagger}U_{x}(k_{x})\hat{\Psi}_{k_{x}}$
of the kicked CL, where $U_{x}(k_{x})$ is given by Eq.~(\ref{eq:Uxkx})
and $\hat{\Psi}_{k_{x}}^{\dagger}\equiv(\hat{u}_{k_{x}}^{\dagger},\hat{v}_{k_{x}}^{\dagger})$,
with $\hat{u}_{k_{x}}^{\dagger}$ ($\hat{v}_{k_{x}}^{\dagger}$) creating
a fermion of quasimomentum $k_{x}$ in the sublattice $u$ ($v$)
pertaining to the upper (lower) leg {[}see Fig.~\ref{fig:Sketch}(a){]}.
Applying the unitary rotation $R=e^{i\frac{\pi}{4}\sigma_{x}}$, we
obtain
\begin{alignat}{1}
	& \widetilde{U}_{x}(k_{x})=RU_{x}(k_{x})R^{\dagger}\nonumber \\
	& = e^{-iJ_{x0}\sigma_{x}}e^{-iJ_{x1}[\cos(k_{x})\sigma_{x}+\sin(k_{x})\sigma_{y}]},\label{eq:UxkxTD}
\end{alignat}
together with the transformed basis
\begin{equation}
	\hat{\Phi}_{k_{x}}\equiv\begin{pmatrix}\hat{a}_{k_{x}}\\
		\hat{b}_{k_{x}}
	\end{pmatrix}=R\begin{pmatrix}\hat{u}_{k_{x}}\\
		\hat{v}_{k_{x}}
	\end{pmatrix}=\frac{\sqrt{2}}{2}\begin{pmatrix}\hat{u}_{k_{x}}+i\hat{v}_{k_{x}}\\
		i\hat{u}_{k_{x}}+\hat{v}_{k_{x}}
	\end{pmatrix}.\label{eq:Phikx}
\end{equation}
In this new basis, the Floquet operator is given by $U_{x}=\sum_{k_{x}}\hat{\Phi}_{k_{x}}^{\dagger}\widetilde{U}_{x}(k_{x})\hat{\Phi}_{k_{x}}$,
and in the lattice representation it takes the form
\begin{equation}
	U_{x}=e^{-iJ_{x0}{\cal H}_{x0}}e^{-iJ_{x1}{\cal H}_{x1}},\label{eq:Ux}
\end{equation}
where
\begin{alignat}{1}
	{\cal H}_{x0} & \equiv\sum_{m}(\hat{a}_{m}^{\dagger}\hat{b}_{m}+{\rm H.c.}),\nonumber \\
	{\cal H}_{x1} & \equiv\sum_{m}(\hat{b}_{m}^{\dagger}\hat{a}_{m+1}+{\rm H.c.}).\label{eq:calU01}
\end{alignat}
The $\hat{a}_{m}^{\dagger}$ and $\hat{b}_{m}^{\dagger}$ are Fourier
transformations of $\hat{a}_{k_{x}}^{\dagger}$ and $\hat{b}_{k_{x}}^{\dagger}$.
To proceed, we solve the Heisenberg equations for $\hat{a}_{m}^{\dagger}$
and $\hat{b}_{m}^{\dagger}$ under the time-evolution governed by
${\cal H}_{x0}$ and ${\cal H}_{x1}$. For a half-infinite ladder
with the unit cell index $m$ going from $1$ to $+\infty$, the solutions
are

\begin{alignat}{1}
	e^{it{\cal H}_{x0}}\hat{a}_{m}^{\dagger}e^{-it{\cal H}_{x0}}= & \cos(t)\hat{a}_{m}^{\dagger}+i\sin(t)\hat{b}_{m}^{\dagger},\nonumber \\
	e^{it{\cal H}_{x0}}\hat{b}_{m}^{\dagger}e^{-it{\cal H}_{x0}}= & \cos(t)\hat{b}_{m}^{\dagger}+i\sin(t)\hat{a}_{m}^{\dagger},\nonumber \\
	e^{it{\cal H}_{x1}}\hat{a}_{m}^{\dagger}e^{-it{\cal H}_{x1}}= & \begin{cases}
		\hat{a}_{m}^{\dagger}, & m=1,\\
		\cos(t)\hat{a}_{m}^{\dagger}+i\sin(t)\hat{b}_{m-1}^{\dagger}, & m>1,
	\end{cases}\nonumber \\
	e^{it{\cal H}_{x1}}\hat{b}_{m}^{\dagger}e^{-it{\cal H}_{x1}}= & \cos(t)\hat{b}_{m}^{\dagger}+i\sin(t)\hat{a}_{m+1}^{\dagger}.
\end{alignat}
Using these solutions, we could further obtain
\begin{widetext}
	\begin{equation}
		U_{x}^{\dagger}\hat{a}_{m}^{\dagger}U_{x}=\begin{cases}
			\cos(J_{x0})\hat{a}_{m}^{\dagger}+i\sin(J_{x0})[\cos(J_{x1})\hat{b}_{m}^{\dagger}+i\sin(J_{x1})\hat{a}_{m+1}^{\dagger}], & m=1,\\
			\cos(J_{x0})[\cos(J_{x1})\hat{a}_{m}^{\dagger}+i\sin(J_{x1})\hat{b}_{m-1}^{\dagger}]+i\sin(J_{x0})[\cos(J_{x1})\hat{b}_{m}^{\dagger}+i\sin(J_{x1})\hat{a}_{m+1}^{\dagger}], & m>1,
		\end{cases}
	\end{equation}
	\begin{equation}
		U_{x}^{\dagger}\hat{b}_{m}^{\dagger}U_{x}=\begin{cases}
			\cos(J_{x0})[\cos(J_{x1})\hat{b}_{m}^{\dagger}+i\sin(J_{x1})\hat{a}_{m+1}^{\dagger}]+i\sin(J_{x0})\hat{a}_{m}^{\dagger}, & m=1,\\
			\cos(J_{x0})[\cos(J_{x1})\hat{b}_{m}^{\dagger}+i\sin(J_{x1})\hat{a}_{m+1}^{\dagger}]+i\sin(J_{x0})[\cos(J_{x1})\hat{a}_{m}^{\dagger}+i\sin(J_{x1})\hat{b}_{m-1}^{\dagger}], & m>1.
		\end{cases}
	\end{equation}
\end{widetext}

Let us denote $|\psi_{{\rm L}}^{(0)}\rangle\equiv\hat{\psi}_{0}^{\dagger}|\emptyset\rangle$
and $|\psi_{{\rm L}}^{(\pi)}\rangle\equiv\hat{\psi}_{\pi}^{\dagger}|\emptyset\rangle$
(``L'' means left) as the eigenstates of $U_{x}$ with quasienergies
$0$ and $\pi$, where $|\emptyset\rangle$ represents the vacuum
state. These zero and $\pi$ eigenmodes should satisfy
\begin{equation}
	U_{x}^{\dagger}\hat{\psi}_{0}^{\dagger}U_{x}=\hat{\psi}_{0}^{\dagger},\qquad U_{x}^{\dagger}\hat{\psi}_{\pi}^{\dagger}U_{x}=-\hat{\psi}_{\pi}^{\dagger}.\label{eq:0pModes}
\end{equation}
To proceed, we expand $|\psi_{{\rm L}}^{(0)}\rangle$ and $|\psi_{{\rm L}}^{(\pi)}\rangle$
in the transformed basis $(\hat{a}_{m}^{\dagger},\hat{b}_{m}^{\dagger})$
as 
\begin{alignat}{1}
	\hat{\psi}_{0}^{\dagger}= & \sum_{m=1}^{\infty}\left(A_{m}\hat{a}_{m}^{\dagger}+B_{m}\hat{b}_{m}^{\dagger}\right),\\
	\hat{\psi}_{\pi}^{\dagger}= & \sum_{m=1}^{\infty}\left(C_{m}\hat{a}_{m}^{\dagger}+D_{m}\hat{b}_{m}^{\dagger}\right).
\end{alignat}
Inserting these expressions into Eq.~(\ref{eq:0pModes}) and solving
the equations for the coefficients $A_{m}$, $B_{m}$,
$C_{m}$ and $D_{m}$, we find that (up to normalization constants):
\begin{alignat}{1}
	\hat{\psi}_{0}^{\dagger}= & \sum_{m=1}^{\infty}\left[-\frac{\tan(J_{x0}/2)}{\tan(J_{x1}/2)}\right]^{m-1}\nonumber\\
	\times& \left[\cos\left(\frac{J_{x0}}{2}\right)\hat{a}_{m}^{\dagger}-i\sin\left(\frac{J_{x0}}{2}\right)\hat{b}_{m}^{\dagger}\right],\\
	\hat{\psi}_{\pi}^{\dagger}= & \sum_{m=1}^{\infty}\left[\frac{1}{\tan(J_{x0}/2)\tan(J_{x1}/2)}\right]^{m-1}\nonumber\\
	\times&\left[\sin\left(\frac{J_{x0}}{2}\right)\hat{a}_{m}^{\dagger}+i\cos\left(\frac{J_{x0}}{2}\right)\hat{b}_{m}^{\dagger}\right].
\end{alignat}
Transforming back to the original basis $(\hat{u}_{m}^{\dagger},\hat{v}_{m}^{\dagger})$
with $\hat{a}_{m}^{\dagger}=\frac{\sqrt{2}}{2}(\hat{u}_{m}^{\dagger}-i\hat{v}_{m}^{\dagger})$
and $\hat{b}_{m}^{\dagger}=\frac{\sqrt{2}}{2}(-i\hat{u}_{m}^{\dagger}+\hat{v}_{m}^{\dagger})$,
we finally obtain the zero and $\pi$ eigenmodes of the kicked CL
as
\begin{alignat}{1}
	|\psi_{{\rm L}}^{(0)}\rangle= & \sum_{m=1}^{\infty}\left[-\frac{\tan(J_{x0}/2)}{\tan(J_{x1}/2)}\right]^{m-1}(\rho_{m}\hat{u}_{m}^{\dagger}-i\lambda_{m}\hat{v}_{m}^{\dagger})|\emptyset\rangle,\label{eq:0SolL}\\
	|\psi_{{\rm L}}^{(\pi)}\rangle= & \sum_{m=1}^{\infty}\left[\frac{1}{\tan(J_{x0}/2)\tan(J_{x1}/2)}\right]^{m-1}\nonumber\\
	\times& (\lambda_{m}\hat{u}_{m}^{\dagger}+i\rho_{m}\hat{v}_{m}^{\dagger})|\emptyset\rangle,\label{eq:pSolL}
\end{alignat}
where $\rho_{m}\equiv\frac{\sqrt{2}}{2}\left[\cos\left(\frac{J_{x0}}{2}\right)-\sin\left(\frac{J_{x0}}{2}\right)\right]$
and $\lambda_{m}\equiv\frac{\sqrt{2}}{2}\left[\cos\left(\frac{J_{x0}}{2}\right)+\sin\left(\frac{J_{x0}}{2}\right)\right]$.
Both the zero and $\pi$ modes should come in pairs due to the degeneracy
enforced by chiral symmetry. The rest of zero and $\pi$ modes
can be obtained by considering the half-infinite ladder with unit-cell
indices $m=-\infty,...,M-1,M$ ($M\gg1$), yielding
\begin{alignat}{1}
	|\psi_{{\rm R}}^{(0)}\rangle= & \sum_{m=-\infty}^{M}\left[-\frac{\tan(J_{x0}/2)}{\tan(J_{x1}/2)}\right]^{M-m}\nonumber\\
	\times&(\lambda_{m}\hat{u}_{m}^{\dagger}+i\rho_{m}\hat{v}_{m}^{\dagger})|\emptyset\rangle,\label{eq:0SolR}\\
	|\psi_{{\rm R}}^{(\pi)}\rangle= & \sum_{m=-\infty}^{M}\left[\frac{1}{\tan(J_{x0}/2)\tan(J_{x1}/2)}\right]^{M-m}\nonumber\\
	\times& (\rho_{m}\hat{u}_{m}^{\dagger}-i\lambda_{m}\hat{v}_{m}^{\dagger})|\emptyset\rangle,\label{eq:pSolR}
\end{alignat}
where ``R'' means right. Since $\rho_{m}$ and $\lambda_{m}$ could
not vanish together, the global profiles of these zero and $\pi$
modes are solely determined by their prefactors $\tan(J_{x0}/2)/\tan(J_{x1}/2)$
and $1/[\tan(J_{x0}/2)\tan(J_{x1}/2)]$.

We now analyze the conditions for these eigenmodes to be edge-localized.
The zero mode in Eq.~(\ref{eq:0SolL}) is exponentially localized
around the left edge ($m=1$) if and only if $|\tan(J_{x1}/2)|>|\tan(J_{x0}/2)|$.
In the phase diagram Fig.~\ref{fig:KCLPD}, this condition is satisfied
in the regions with $J_{x1}\in(J_{x0},2\pi-J_{x0})$ for $J_{x0}\in(0,\pi)$
and with $J_{x1}\in(2\pi-J_{x0},J_{x0})$ for $J_{x0}\in(\pi,2\pi)$.
The $\pi$ mode in Eq.~(\ref{eq:pSolL}) is exponentially localized
around the left edge ($m=1$) if and only if $|\tan(J_{x0}/2)\tan(J_{x1}/2)|>1$.
In the phase diagram Fig.~\ref{fig:KCLPD}, this condition is fulfilled
in the regions with $J_{x1}\in(\pi-J_{x0},\pi+J_{x0})$ for $J_{x0}\in(0,\pi)$
and with $J_{x1}\in(J_{x0}-\pi,3\pi-J_{x0})$ for $J_{x0}\in(\pi,2\pi)$.
Referring to the winding numbers $(w_{0x},w_{\pi x})$ in Fig.~\ref{fig:KCLPD},
their relationship with the invariants $(\omega_{0x},\omega_{\pi x})$
analyzed below Eq.~(\ref{eq:ome0px}), and the chiral-symmetry-enforced
twofold degeneracy of zero and $\pi$ eigenmodes, we end up with the
bulk-edge correspondence for the kicked CL as
\begin{equation}
	(N_{0x},N_{\pi x})=2(|w_{0x}|,|w_{\pi x}|)=2(|\omega_{0x}|,|\omega_{\pi x}|),\label{eq:KCLBBC1}
\end{equation}
where $N_{0x}$ ($N_{\pi x}$) counts the number of exponentially
localized edge modes at quasienergy zero ($\pi$). This relation holds
true among the gapped phases of the kicked CL. 

We next consider the bulk-edge correspondence at phase boundaries.
Along the critical lines $J_{x1}=\pi\pm J_{x0}$ with $J_{x0}\in(0,\pi)$,
$J_{x1}=-\pi+J_{x0}$ with $J_{x0}\in(\pi,2\pi)$, and $J_{x1}=3\pi-J_{x0}$
with $J_{x0}\in(\pi,2\pi)$, our solution in Eq.~(\ref{eq:0SolL})
permits the survival of one localized Floquet zero mode at each edge
{[}$(N_{0x},N_{\pi x})=(2,0)${]}. The generalized winding numbers
in Eq.~(\ref{eq:ome0px}) take $(\omega_{0x},\omega_{\pi x})=(1,0)$
along these phase boundaries, while the conventional winding numbers
in Eq.~(\ref{eq:w0px}) yield $(w_{0x},w_{\pi x})=(1,1/2)$. Along
the critical lines $J_{x1}=J_{x0}$ and $J_{x1}=2\pi-J_{x0}$ with
$J_{x0}\in(\pi/2,3\pi/2)$, our solution in Eq.~(\ref{eq:pSolL})
permits the survival of one localized Floquet $\pi$ mode at each
edge {[}$(N_{0x},N_{\pi x})=(0,2)${]}. The generalized winding numbers
in Eq.~(\ref{eq:ome0px}) take $(\omega_{0x},\omega_{\pi x})=(0,1)$
along these phase boundaries, while the conventional winding numbers
in Eq.~(\ref{eq:w0px}) yield $(w_{0x},w_{\pi x})=(1/2,1)$. Along
the other critical lines in Fig.~\ref{fig:KCLPD}, the solutions in
Eqs.~(\ref{eq:0SolL}) and (\ref{eq:pSolL}) are no longer Floquet
edge modes. The generalized winding numbers in Eq.~(\ref{eq:ome0px})
take $(\omega_{0x},\omega_{\pi x})=(0,0)$ along these phase boundaries,
while at least one of the conventional winding numbers $(w_{0x},w_{\pi x})$
in Eq.~(\ref{eq:w0px}) takes a half-integer value. 

Overall, we find that the invariants $(\omega_{0x},\omega_{\pi x})$
could always count the numbers of Floquet zero and $\pi$ edge modes
along the critical lines correctly, in the sense that $(N_{0x},N_{\pi x})=2(|\omega_{0x}|,|\omega_{\pi x}|)$,
whereas the conventional winding numbers $(w_{0x},w_{\pi x})$ may
become invalid due to the emergence of half-quantized values. Looking
back at Eq.~(\ref{eq:KCLBBC1}), we finally conclude that Eq.~(\ref{eq:KCLBBC})
in the main text describes the most general bulk-edge correspondence
of the kicked CL, working equally well in its gapped phases and along
its phase boundaries where the bulk spectra become gapless. We expect
the Eq.~(\ref{eq:KCLBBC}) to hold also in other 1D, two-band chiral-symmetric
driven systems, regardless of the presence of a quasienergy gap.

\section{Floquet zero and $\pi$ corner modes of the periodically
kicked CL-SSH model}\label{sec:App3}

In Appendices \ref{sec:App1} and \ref{sec:App2}, we have analytically
obtained the zero-energy edge modes of the SSH model and the ${\rm zero}/\pi$-quasienergy
edge modes of the periodically kicked CL. Since the Floquet operator
of our 2D driven CL-SSH model is given by 
\begin{equation}
	U=U_{x}\otimes U_{y},\label{eq:Uapp}
\end{equation}
with $U_{x}=e^{-iJ_{x0}{\cal H}_{x0}}e^{-iJ_{x1}{\cal H}_{x1}}$ 
{[}Eq.~(\ref{eq:Ux}){]} and $U_{y}=e^{-iH_{y}}$ {[}Eq.~(\ref{eq:Hy}){]},
the corner-localized solutions of $U$ at zero and $\pi$ quasienergies
can be deduced from the edge modes of $U_{x}$ and $U_{y}$ under
OBCs.

First, we notice that any Floquet eigenstate of $U$ must be in the
direct product form $|\Psi\rangle=|\psi\rangle\otimes|\psi'\rangle$
with quasienergy ${\cal E}=\varepsilon+\varepsilon'$ mod $2\pi$,
where $|\psi\rangle$ ($|\psi'\rangle$) is an eigenstate of $U_{x}$
($U_{y}$) with quasienergy $\varepsilon$ ($\varepsilon'$). Second,
for the $|\Psi\rangle$ to be a corner-localized eigenmode, neither
$|\psi\rangle$ nor $|\psi'\rangle$ should be an extended bulk state.
Otherwise, $|\Psi\rangle$ would be extended along at least one spatial
dimension. Therefore, $|\Psi\rangle=|\psi\rangle\otimes|\psi'\rangle$
is corner-localized if and only if both $|\psi\rangle$ and $|\psi'\rangle$
are localized eigenstates of their corresponding propagators $U_{x}$
and $U_{y}$ under OBCs. Third, the edge-state solutions of $U_{y}$
are also those of $H_{y}$ {[}Eq.~(\ref{eq:Hy}){]}, whose quasienergy
must be zero {[}Eqs.~(\ref{eq:0y1}) and (\ref{eq:0y2}){]}. Meanwhile,
the edge-state solutions of $U_{x}$ {[}Eq.~(\ref{eq:Ux}){]} have
quasienergy zero {[}Eqs.~(\ref{eq:0SolL}) and (\ref{eq:0SolR}){]}
or $\pi$ {[}Eqs.~(\ref{eq:pSolL}) and (\ref{eq:pSolR}){]}. Therefore,
the corner-localized solutions of $U=U_{x}\otimes U_{y}$ could only
appear at the quasienergy ${\cal E}=0$ or ${\cal E}=\pm\pi$.

With these considerations, we can find all the possibly corner-localized
eigenmodes of $U$ under the OBCs as
\begin{alignat}{1}
& |\Psi_{{\rm LB}}^{({\cal E})}\rangle=|\psi_{{\rm L}}^{({\cal E})}\rangle\otimes|\psi'_{{\rm B}}\rangle,\quad|\Psi_{{\rm LT}}^{({\cal E})}\rangle=|\psi_{{\rm L}}^{({\cal E})}\rangle\otimes|\psi'_{{\rm T}}\rangle,\label{eq:FCM1}\\
& |\Psi_{{\rm RB}}^{({\cal E})}\rangle=|\psi_{{\rm R}}^{({\cal E})}\rangle\otimes|\psi'_{{\rm B}}\rangle,\quad|\Psi_{{\rm RT}}^{({\cal E})}\rangle=|\psi_{{\rm R}}^{({\cal E})}\rangle\otimes|\psi'_{{\rm T}}\rangle,\label{eq:FCM2}
\end{alignat}
where ${\cal E}=0$ and $\pi$. Explicit expressions of $|\psi_{{\rm L,R}}^{(0)}\rangle$,
$|\psi_{{\rm L,R}}^{(\pi)}\rangle$ and $|\psi'_{{\rm B,T}}\rangle$
are given by Eqs.~(\ref{eq:0SolL})--(\ref{eq:pSolR})
and (\ref{eq:0y1})--(\ref{eq:0y2}). It is clear that
once these eigenstates appear as corner modes with the quasienergy
${\cal E}$ ($0$ or $\pi$), they must be fourfold degenerate. From
the wave functions in Eqs.~(\ref{eq:FCM1}) and (\ref{eq:FCM2}),
we can further identify the conditions for these states to be corner-localized,
as detailed below.

(i) $|\psi_{{\rm L,R}}^{(0)}\rangle$ are edge-localized zero modes
if and only if $|\tan(J_{x1}/2)|>|\tan(J_{x0}/2)|$ {[}Eqs.~(\ref{eq:0SolL})
and (\ref{eq:0SolR}){]}.

(ii) $|\psi_{{\rm L,R}}^{(\pi)}\rangle$ are edge-localized $\pi$
modes if and only if $|\tan(J_{x0}/2)\tan(J_{x1}/2)|>1$ {[}Eqs.~(\ref{eq:pSolL})
and (\ref{eq:pSolR}){]}.

(iii) $|\psi'_{{\rm B,T}}\rangle$ are edge-localized zero modes if
and only if $|J_{y1}|>|J_{y0}|$ {[}Eqs.~(\ref{eq:0y1}) and (\ref{eq:0y2}){]}.

(iv) Referring to (i) and (iii), the states $|\Psi_{{\rm LB}}^{(0)}\rangle$,
$|\Psi_{{\rm LT}}^{(0)}\rangle$, $|\Psi_{{\rm RB}}^{(0)}\rangle$
and $|\Psi_{{\rm RT}}^{(0)}\rangle$ constitute four corner-localized
Floquet eigenmodes with quasienergy ${\cal E}=0$ if and only if
\begin{equation}
	|\tan(J_{x1}/2)|>|\tan(J_{x0}/2)|\quad\&\quad|J_{y1}|>|J_{y0}|.\label{eq:FC0M}
\end{equation}

(v) Referring to (ii) and (iii), the states $|\Psi_{{\rm LB}}^{(\pi)}\rangle$,
$|\Psi_{{\rm LT}}^{(\pi)}\rangle$, $|\Psi_{{\rm RB}}^{(\pi)}\rangle$
and $|\Psi_{{\rm RT}}^{(\pi)}\rangle$ constitute four corner-localized
Floquet eigenmodes with quasienergy ${\cal E}=\pi$ if and only if
\begin{equation}
	|\tan(J_{x0}/2)\tan(J_{x1}/2)|>1\quad\&\quad|J_{y1}|>|J_{y0}|.\label{eq:FCpM}
\end{equation}

(vi) There will be four Floquet corner modes at ${\cal E}=0$ and
four other corner modes at ${\cal E}=\pi$ if and only if the conditions
in Eqs.~(\ref{eq:FC0M}) and (\ref{eq:FCpM}) are both satisfied.

(vii) When none of the conditions in Eqs.~(\ref{eq:FC0M}) and (\ref{eq:FCpM})
are fulfilled, the system will have no corner modes.

These analytical results are consistent with our numerical calculations,
especially along phase boundaries between different SOTPs as reported
in the main text.

\section{Bulk-corner correspondence of the periodically kicked
	CL-SSH model}\label{sec:App4}
In this Appendix, we propose the topological invariants that could
count the numbers of zero and $\pi$ Floquet corner modes in each
parameter region, thereby establishing the rule of bulk-corner correspondence
for our kicked CL-SSH model. Importantly, this rule allows us to know
along which phase boundary the zero or $\pi$ Floquet corner modes
could survive topological phase transitions, making the associated
critical points topologically nontrivial. It also applies to Floquet
SOTPs in other 2D systems with chiral symmetry and under similar driving
protocols (i.e., only system parameters along one of the two spatial
dimensions depend on driving fields).

In the main text, we have introduced the generalized invariant
$\omega_{y}$ {[}Eq.~(\ref{eq:omey}){]} and established the bulk-edge
correspondence {[}Eq.~(\ref{eq:SSHBBC}){]} for the static SSH model.
The Eq.~(\ref{eq:omey}) can also describe the topological phases
and bulk-edge correspondence of the one-period free-evolution operator
$U_{y}=e^{-iH_{y}}$ of the SSH model {[}with the $H_{y}$ given by
Eq.~(\ref{eq:Hy}){]}. On the other hand, the bulk-edge correspondence
of the kicked CL is captured by the generalized invariants
$(\omega_{0x},\omega_{\pi x})$ in Eq.~(\ref{eq:KCLBBC1}). Since
the zero and $\pi$ corner modes of the driven CL-SSH model
{[}Eq.~(\ref{eq:Hkt}){]} can only be formed via coupling edge states
of its parent systems, we can infer the bulk-corner correspondence
of the kicked CL-SSH model as
\begin{equation}
	(N_{0},N_{\pi})=4(|\omega_{0x}\omega_{y}|,|\omega_{\pi x}\omega_{y}|)\equiv4(\omega_{0},\omega_{\pi}).\label{eq:CLSSHBCC1}
\end{equation}
The invariants $(\omega_{0},\omega_{\pi})$, as defined in Eq.~(\ref{eq:CLSSHBCC1}),
could characterize the higher-order topology and bulk-corner correspondence
of all the gapped and gapless phases in our kicked CL-SSH model.

We can verify Eq.~(\ref{eq:CLSSHBCC1}) as follows. First, due to
the Eq.~(\ref{eq:SSHBBC}), we only need to prove the following relations
for the kicked CL-SSH model
\begin{equation}
	(N_{0},N_{\pi})=\begin{cases}
		(0,0), & |J_{y1}|\leq|J_{y0}|,\\
		4(|\omega_{0x}|,|\omega_{\pi x}|), & |J_{y1}|>|J_{y0}|.
	\end{cases}\label{eq:CLSSHBCC2}
\end{equation}
According to Eqs.~(\ref{eq:FC0M}) and (\ref{eq:FCpM}), we will have
no zero or $\pi$ Floquet corner modes when $|J_{y1}|\leq|J_{y0}|$,
i.e., $(N_{0},N_{\pi})=(0,0)$. The first equality in Eq.~(\ref{eq:CLSSHBCC2})
is thus confirmed. Under the condition $|J_{y1}|>|J_{y0}|$, we can
further analyze the values of $\omega_{0x}$ and $\omega_{\pi x}$
following the Appendix \ref{sec:App2} and Eq.~(\ref{eq:PB}).

To proceed, we factorize the complex-continued functions in Eq.~(\ref{eq:f12xz})
as follows
\begin{widetext}
	\begin{alignat}{1}
		f_{1x}(z)= & -\sin J_{x0}\sin^{2}\frac{J_{x1}}{2}\left(z+\frac{\tan\frac{J_{x0}}{2}}{\tan\frac{J_{x1}}{2}}\right)\left(z-\frac{1}{\tan\frac{J_{x0}}{2}\tan\frac{J_{x1}}{2}}\right),\nonumber \\
		f_{2x}(z)= & \frac{\sin J_{x1}\cos^{2}\frac{J_{x0}}{2}\left(z+\frac{\tan\frac{J_{x0}}{2}}{\tan\frac{J_{x1}}{2}}\right)\left(z-\tan\frac{J_{x0}}{2}\tan\frac{J_{x1}}{2}\right)z}{z^{2}}.\label{eq:f12xz2}
	\end{alignat}
\end{widetext}
The function $f_{1x}(z)$ has no poles. Referring to Eq.~(\ref{eq:ome12x}),
we should have
\begin{widetext}
	\begin{equation}
		\omega_{1x}=\begin{cases}
			2, & \left|\tan\frac{J_{x1}}{2}\right|>\left|\tan\frac{J_{x0}}{2}\right|\quad\&\quad\left|\tan\frac{J_{x0}}{2}\tan\frac{J_{x1}}{2}\right|>1,\\
			1, & \left|\tan\frac{J_{x1}}{2}\right|>\left|\tan\frac{J_{x0}}{2}\right|\quad\&\quad\left|\tan\frac{J_{x0}}{2}\tan\frac{J_{x1}}{2}\right|\leq1,\\
			1, & \left|\tan\frac{J_{x1}}{2}\right|\le\left|\tan\frac{J_{x0}}{2}\right|\quad\&\quad\left|\tan\frac{J_{x0}}{2}\tan\frac{J_{x1}}{2}\right|>1,\\
			0, & \left|\tan\frac{J_{x1}}{2}\right|\le\left|\tan\frac{J_{x0}}{2}\right|\quad\&\quad\left|\tan\frac{J_{x0}}{2}\tan\frac{J_{x1}}{2}\right|\le1.
		\end{cases}\label{eq:ome1x}
	\end{equation}
\end{widetext}
Next, referring to the Eq.~(\ref{eq:PB}), it is straightforward to
verify that the conditions $J_{x0}\pm J_{x1}=(2\nu+1)\pi$ and $\left|\tan\frac{J_{x0}}{2}\tan\frac{J_{x1}}{2}\right|\neq1$
cannot be simultaneously fulfilled. Therefore, we can use $\omega_{2x}=N_{2z}-N_{2p}$
to find the $\omega_{2x}$ from $f_{2x}(z)$ when $\left|\tan\frac{J_{x0}}{2}\tan\frac{J_{x1}}{2}\right|\neq1$,
yielding
\begin{widetext}
	\begin{equation}
		\omega_{2x}=\begin{cases}
			0, & \left|\tan\frac{J_{x1}}{2}\right|>\left|\tan\frac{J_{x0}}{2}\right|\quad\&\quad\left|\tan\frac{J_{x0}}{2}\tan\frac{J_{x1}}{2}\right|>1,\\
			1, & \left|\tan\frac{J_{x1}}{2}\right|>\left|\tan\frac{J_{x0}}{2}\right|\quad\&\quad\left|\tan\frac{J_{x0}}{2}\tan\frac{J_{x1}}{2}\right|<1,\\
			-1, & \left|\tan\frac{J_{x1}}{2}\right|\le\left|\tan\frac{J_{x0}}{2}\right|\quad\&\quad\left|\tan\frac{J_{x0}}{2}\tan\frac{J_{x1}}{2}\right|>1,\\
			0, & \left|\tan\frac{J_{x1}}{2}\right|\le\left|\tan\frac{J_{x0}}{2}\right|\quad\&\quad\left|\tan\frac{J_{x0}}{2}\tan\frac{J_{x1}}{2}\right|<1.
		\end{cases}\label{eq:ome2xint}
	\end{equation}
\end{widetext}
When $J_{x0}\pm J_{x1}=(2\nu+1)\pi$, we have $\Delta_{\pi}=0$ and
$\left|\tan\frac{J_{x0}}{2}\tan\frac{J_{x1}}{2}\right|=1$. According
to Eq.~(\ref{eq:ome12x}), we should use $\omega_{2x}=N_{2z}-N_{2p}/2$
to compute the $\omega_{2x}$ from $f_{2x}(z)$ in this case, yielding
$\omega_{2x}=1$ for $\left|\tan\frac{J_{x1}}{2}\right|>\left|\tan\frac{J_{x0}}{2}\right|$
and $\omega_{2x}=0$ for $\left|\tan\frac{J_{x1}}{2}\right|\leq\left|\tan\frac{J_{x0}}{2}\right|$.
These results can be combined into the second and fourth lines of
Eq.~(\ref{eq:ome2xint}), leaving us with
\begin{widetext}
	\begin{equation}
		\omega_{2x}=\begin{cases}
			0, & \left|\tan\frac{J_{x1}}{2}\right|>\left|\tan\frac{J_{x0}}{2}\right|\quad\&\quad\left|\tan\frac{J_{x0}}{2}\tan\frac{J_{x1}}{2}\right|>1,\\
			1, & \left|\tan\frac{J_{x1}}{2}\right|>\left|\tan\frac{J_{x0}}{2}\right|\quad\&\quad\left|\tan\frac{J_{x0}}{2}\tan\frac{J_{x1}}{2}\right|\leq1,\\
			-1, & \left|\tan\frac{J_{x1}}{2}\right|\le\left|\tan\frac{J_{x0}}{2}\right|\quad\&\quad\left|\tan\frac{J_{x0}}{2}\tan\frac{J_{x1}}{2}\right|>1,\\
			0, & \left|\tan\frac{J_{x1}}{2}\right|\le\left|\tan\frac{J_{x0}}{2}\right|\quad\&\quad\left|\tan\frac{J_{x0}}{2}\tan\frac{J_{x1}}{2}\right|\leq1.
		\end{cases}\label{eq:ome2x}
	\end{equation}
\end{widetext}
Plugging Eqs.~(\ref{eq:ome1x}) and (\ref{eq:ome2x}) into the definition
of $(\omega_{0x},\omega_{\pi x})$ in
Eq.~(\ref{eq:ome2x}) finally yields
\begin{widetext}
	\begin{equation}
		(\omega_{0x},\omega_{\pi x})=\begin{cases}
			(1,1), & \left|\tan\frac{J_{x1}}{2}\right|>\left|\tan\frac{J_{x0}}{2}\right|\quad\&\quad\left|\tan\frac{J_{x0}}{2}\tan\frac{J_{x1}}{2}\right|>1,\\
			(1,0), & \left|\tan\frac{J_{x1}}{2}\right|>\left|\tan\frac{J_{x0}}{2}\right|\quad\&\quad\left|\tan\frac{J_{x0}}{2}\tan\frac{J_{x1}}{2}\right|\leq1,\\
			(0,1), & \left|\tan\frac{J_{x1}}{2}\right|\le\left|\tan\frac{J_{x0}}{2}\right|\quad\&\quad\left|\tan\frac{J_{x0}}{2}\tan\frac{J_{x1}}{2}\right|>1,\\
			(0,0), & \left|\tan\frac{J_{x1}}{2}\right|\le\left|\tan\frac{J_{x0}}{2}\right|\quad\&\quad\left|\tan\frac{J_{x0}}{2}\tan\frac{J_{x1}}{2}\right|\leq1.
		\end{cases}\label{eq:ome0pxKCL}
	\end{equation}
\end{widetext}
If the rule of bulk-corner correspondence in Eq.~(\ref{eq:CLSSHBCC1})
is true, we should have the following results for our kicked CL-SSH
model according to Eqs.~(\ref{eq:CLSSHBCC2}) and (\ref{eq:ome0pxKCL}),
i.e.,
\begin{widetext}
	\begin{equation}
		(N_{0},N_{\pi})=\begin{cases}
			(4,4), & |J_{y1}|>|J_{y0}|\quad\&\quad\left|\tan\frac{J_{x1}}{2}\right|>\left|\tan\frac{J_{x0}}{2}\right|\quad\&\quad\left|\tan\frac{J_{x0}}{2}\tan\frac{J_{x1}}{2}\right|>1,\\
			(4,0), & |J_{y1}|>|J_{y0}|\quad\&\quad\left|\tan\frac{J_{x1}}{2}\right|>\left|\tan\frac{J_{x0}}{2}\right|\quad\&\quad\left|\tan\frac{J_{x0}}{2}\tan\frac{J_{x1}}{2}\right|\leq1,\\
			(0,4), & |J_{y1}|>|J_{y0}|\quad\&\quad\left|\tan\frac{J_{x1}}{2}\right|\le\left|\tan\frac{J_{x0}}{2}\right|\quad\&\quad\left|\tan\frac{J_{x0}}{2}\tan\frac{J_{x1}}{2}\right|>1,\\
			(0,0), & {\rm Otherwise}.
		\end{cases}\label{eq:BCCfin}
	\end{equation}
\end{widetext}
These relations are consistent with our counting of zero
and $\pi$ corner modes $N_{0}$ and $N_{\pi}$ following
the points (i)--(vii) of Appendix \ref{sec:App3}, which are nevertheless
deduced from exact solutions of corner states. The proposed rule of
bulk-corner correspondence in Eq.~(\ref{eq:CLSSHBCC1}) is then verified.

\begin{figure*}
	\begin{centering}
		\includegraphics[scale=0.37]{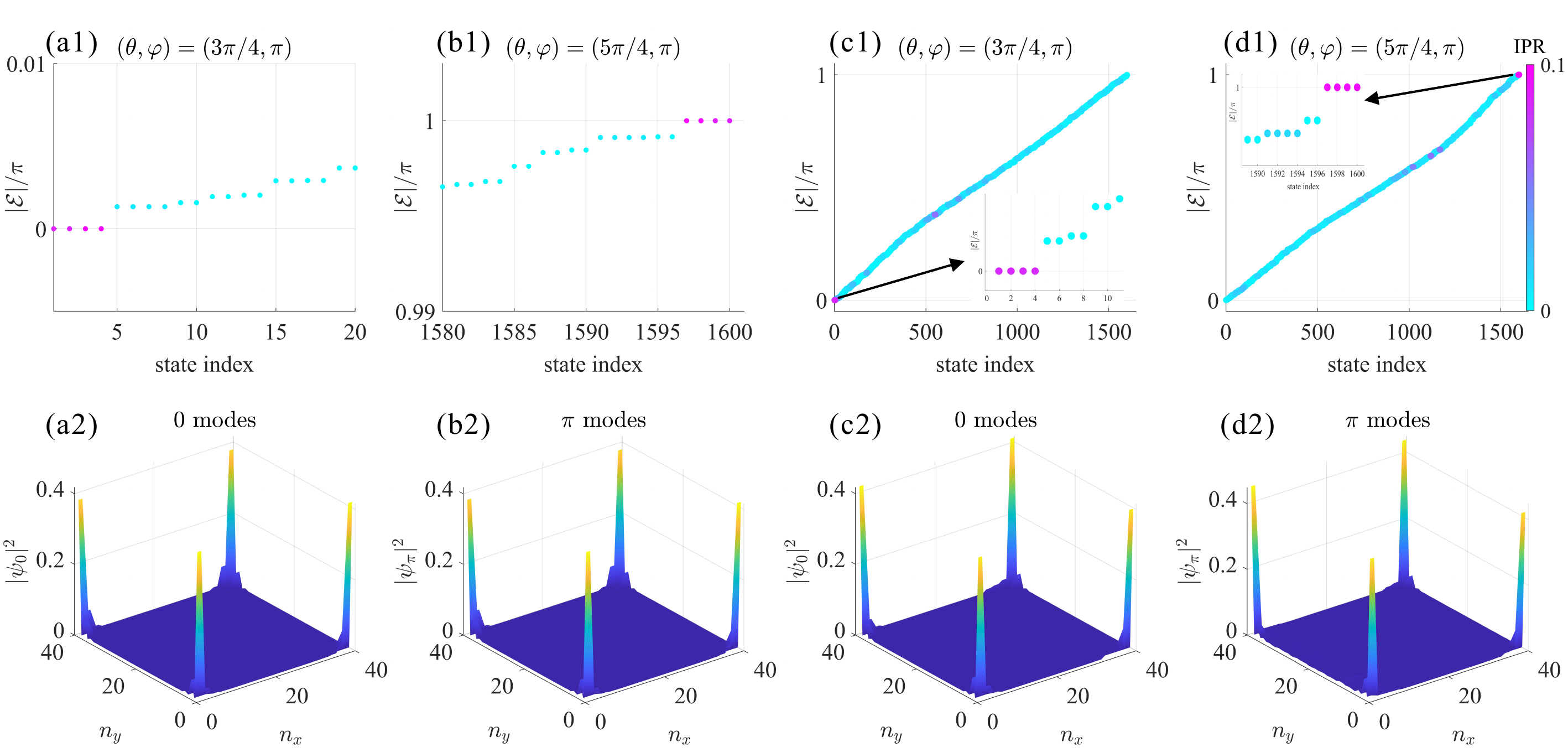}
		\par\end{centering}
	\caption{Floquet spectra and corner modes of the kicked CL-SSH model at the
		critical points $(\theta,\varphi)=(3\pi/4,\pi)$ {[}for (a1), (a2),
		(c1), (c2){]} and $(5\pi/4,\pi)$ {[}for (b1), (b2), (d1), (d2){]}
		under perturbations. The perturbation parameters are $\delta_{x}=\delta_{y}=0.1$
		and $\delta_{1}=\delta_{2}=\lambda=0.2$. (a1) Absolute values of
		the first $20$ quasienergies and their corresponding IPRs, with perturbations
		given in Eqs.~(\ref{eq:H1kxyp}) and (\ref{eq:H2kxyp}). (a2) Probability
		distributions of the four zero modes in (a1). (b1) Absolute values
		of the last $20$ quasienergies and their corresponding IPRs, with
		perturbations given in Eqs.~(\ref{eq:H1kxyp}) and (\ref{eq:H2kxyp}).
		(b2) Probability distributions of the four $\pi$ modes in (b1). (c1)
		Absolute values of quasienergies and their corresponding IPRs with
		hopping disorder. The first $11$ of them are shown in the inner panel.
		(c2) Probability distributions of the four zero modes in (c1). (d1)
		Absolute values of quasienergies and their corresponding IPRs with
		hopping disorder. The last $12$ of them are shown in the inner panel.
		(d2) Probability distributions of the four $\pi$ modes in (d1).\label{fig:DSOD}}
\end{figure*}

\section{Robustness to perturbations}\label{sec:App5}

In this Appendix, we reveal the robustness of Floquet corner modes
at the critical points to perturbations that break the tensor product
structure and the symmetries except the chiral
symmetry of the Floquet operator. We focus on the kicked CL-SSH model to demonstrate our results.

We first introduce perturbations to break the symmetries and tensor
product structure of our Floquet operator. In the momentum space,
the perturbed Floquet operator has the form $U(k_{x},k_{y})=e^{-iH_{1}(k_{x},k_{y})}e^{-iH_{2}(k_{x},k_{y})}$,
\begin{alignat}{1}
	H_{1}(k_{x},k_{y})& = (J_{x0}\sigma_{x}-\delta_{x}\cos k_{x}\sigma_{z})\otimes\tau_{0}\nonumber \\
	& +\delta_{1}\cos k_{x}\sigma_{x}\otimes\tau_{z}\nonumber \\
	& + \sigma_{0}\otimes[H_{y}(k_{y})+\delta_{y}\cos k_{y}\tau_{y}],\label{eq:H1kxyp}
\end{alignat}
\begin{alignat}{1}
	H_{2}(k_{x},k_{y})& = J_{x1}(\cos k_{x}\sigma_{x}+\sin k_{x}\sigma_{z})\otimes\tau_{0}\nonumber \\
	& +\delta_{2}\sin k_{x}\sigma_{z}\otimes\tau_{z}\nonumber \\
	& + \sigma_{0}\otimes[H_{y}(k_{y})+\delta_{y}\cos k_{y}\tau_{y}],\label{eq:H2kxyp}
\end{alignat}
where $H_{y}(k_{y})$ is the Bloch Hamiltonian of the static SSH model.
The perturbations proportional to $\delta_{x}$ and $\delta_{y}$
break the time-reversal, particle-hole and inversion symmetries of
the system. Despite the rectangular lattice geometry and the chiral
symmetry ${\cal S}_{x}\otimes{\cal S}_{y}=\sigma_{y}\otimes\tau_{z}$,
other lattice symmetries of the system are all broken. The terms proportional
to $\delta_{1}$ and $\delta_{2}$ are noncommuting and further destroy
the factorization structure of the original Floquet operator $U(k_{x},k_{y})=U(k_{x})\otimes U(k_{y})$.
With all these perturbations, we illustrate in Figs.~\ref{fig:DSOD}(a1),
\ref{fig:DSOD}(a2), \ref{fig:DSOD}(b1), and \ref{fig:DSOD}(b2)
that the Floquet zero/$\pi$ topological corner modes and their corner
localization are all retained. Therefore, our found gapless higher-order
Floquet topology is robust to perturbations breaking the lattice symmetries
and the tensor product structure of the Floquet operator. The chiral
symmetry is the only essential symmetry that protecting the topological
corner modes in our system.

To further check the robustness of critical Floquet corner modes to
perturbations, we add disorder $\mu_{m,n}=\lambda\epsilon_{m,n}$
to all the nearest-neighbor hopping terms of our kicked CL-SSH model,
where $m$ and $n$ denote lattice indices along $x$ and $y$ directions.
Each $\epsilon_{m,n}$ is taken randomly in the range $[-1,1]$. Despite
breaking the translational symmetry, these random hopping terms also
break the other lattice symmetries and the tensor product structure
of the original Floquet model, keeping only the chiral symmetry untouched.
In Figs.~\ref{fig:DSOD}(c1), \ref{fig:DSOD}(d1), \ref{fig:DSOD}(c2),
and \ref{fig:DSOD}(d2), we show the quasienergy spectra and corner
modes of the disordered Floquet system, with other system parameters
taken at two distinct critical points of the original clean model.
We considered $50$ disorder realizations and randomly picked two
of the Floquet spectra following different disorder patterns for each case to show
in Figs.~\ref{fig:DSOD}(c1) and \ref{fig:DSOD}(d1). The results
clearly indicate that both the Floquet zero/$\pi$ corner modes and
their corner localization are maintained in the presence of disorder.
Therefore, our found gapless higher-order Floquet topology is robust
to chiral-symmetry-preserving disorder, even though the bulk topological
numbers of the clean system cannot be directly computed in disordered
cases.

\end{document}